\documentclass[12pt, fleqn, a4paper]{article}

\usepackage[utf8]{inputenc}
\usepackage[american]{babel}
\usepackage{amsmath, amssymb, amsthm}

\usepackage[backend=biber, style=authoryear, uniquename=false, uniquelist=false]{biblatex}
\usepackage{geometry, setspace}
\AtBeginEnvironment{tabular}{\doublespacing} 
\AtEndEnvironment{tabular}{\vspace{6pt}}
\AtBeginEnvironment{longtable}{\doublespacing} 
\AtEndEnvironment{longtable}{\vspace{6pt}}

\usepackage{appendix}

\usepackage{graphicx, subfigure} 
\usepackage{booktabs, longtable} 
\usepackage{caption, placeins} 
\usepackage{hyperref} 
\hypersetup{
    colorlinks=true,
    urlcolor=blue,
    linkcolor=red,
    citecolor=blue
}
\usepackage[nameinlink]{cleveref} 

\crefname{figure}{figure}{figures}
\Crefname{figure}{Figure}{Figures}
\crefname{cTable}{table}{tables}
\Crefname{CTable}{Table}{Tables}
\crefname{equation}{equation}{equations}
\Crefname{equation}{Equation}{Equations}

\usepackage{orcidlink} 
\usepackage{authblk} 

\begin{document}	 
	
	\title{
		
		\vspace{-2cm} 
		
		Impact of Tobacco Advertising Restrictions in Switzerland: A Quasi-Experimental Study on the Effect of Billboard Bans on Smoking\thanks{
			Corresponding author: Andreas Stoller, University of Fribourg, Department of Economics, Bd de P{\'e}rolles 90, 1700 Fribourg, Switzerland, andreas.stoller@unifr.ch
			}
		
		}
	
	\date{August 25, 2026 \vspace{-1cm}}
	\author[]{Andreas Stoller$^{1}$ \orcidlink{0000-0002-2404-648X} \vspace{-0.5cm}}
	\affil[]{$^{1}$University of Fribourg, Department of Economics \vspace{-0.5cm}}
	
	\maketitle
	
	\begin{abstract}
		
		This study assesses the impact of tobacco billboard bans on smoking in Switzerland, exploiting their staggered adoption across regions, i.e., the cantons. Based on retrospective smoking histories from the Swiss Health Survey, a panel of individuals’ annual smoking status is reconstructed, containing more than one million observations from 1993 to 2017. Estimation relies on staggered difference-in-differences as well as a complementary latent factor model, which relaxes the common trends assumption. The findings indicate that tobacco billboard bans lead to a reduction in smoking rates. Reductions of up to 0.9 percentage points correspond to an approximate 3\% decline in the smoking rate. The effect is driven by women and individuals aged 25--44 and 65+. Overall, this evidence suggests that even partial tobacco advertising bans, such as billboard bans, can effectively reduce smoking rates and serve as a valuable policy tool within comprehensive tobacco prevention strategies.
		
		\begin{flushleft}
			
			
			
		\end{flushleft}
		
	\end{abstract}

\newpage

\section{Introduction}

Tobacco smoking accounted for 8.71 million deaths globally in 2019, representing 15.4\% of all deaths \parencite{Murray20}. Despite the remarkable decline in smoking over the last decades, smoking remains one of the leading risk factors \parencite{Reitsma21}. While the rise of new products such as e-cigarettes is concerning, the regulation of traditional smoking remains a key challenge for public health. Despite the extensive evaluation of tobacco prevention policies, certain policy tools remain relatively understudied. Numerous studies focus on tobacco price and tax policies to reduce smoking, suggesting their effectiveness \parencite{DeCicca22}. Place-based smoking restrictions, such as indoor smoking bans, also appear promising, particularly in Europe. However, the evidence on other non-price policies, such as advertising restrictions, is less conclusive and warrants further investigation.

Up until 2000, the role of tobacco advertising in smoking was extensively studied \parencite{Saffer00, Gallet03, Goel06}. Tobacco advertising may initiate smoking among non-smokers or increase consumption among smokers, thereby expanding overall market size. However, it may also merely shift smokers across brands without increasing overall demand. Consequently, its relevance for tobacco prevention is not straightforward. While some empirical evidence supports the market expansion hypothesis, other studies do not. Existing analyses face considerable data and methodological limitations, which could be addressed through recent advances in quasi-experimental methods and access to more granular data.

Switzerland presents a substantial variation in tobacco advertising restrictions across its geographic regions, i.e., the cantons \parencite{BAG25cantonal}. Beginning in 1997, an increasing number of cantons introduced tobacco billboard bans, with a majority implementing such restrictions by 2017. At the time, billboard advertising was the most important expenditure for the tobacco industry, making these bans particularly relevant \parencite{BAG15}. The quasi-experimental setting allows us to exploit the variation in tobacco billboard bans across time and cantons to assess their impact on smoking. 

This study assesses the effect of tobacco billboard bans on smoking in Switzerland. The analysis relies on two complementary methods, staggered difference-in-differences (staggered DiD) and interactive fixed effects counterfactual (IFEct) estimation, both robust to treatment effect heterogeneity \parencite{Callaway21, Liu24}. The IFEct estimator suggests an immediate and sustained reduction in the smoking rate by 0.9 percentage points (pp), with a p-value $<$ 0.01, statistically significant at the 5\% level. The staggered DiD results are less pronounced, averaging a reduction of 0.4 pp (p $=$ 0.08) over five years after the tobacco billboard ban adoption. Both approaches indicate that the tobacco billboard bans lead to a reduction in the smoking rate, in line with the market expansion hypothesis.

This paper aims to contribute to the literature on the effectiveness of tobacco advertising bans. The focus lies on their effect on smoking, and ultimately, their role as a policy tool to improve public health. The analysis is based on a considerably large individual-level dataset. This improves upon previous studies, which typically relied on aggregated data at the country level and faced related endogeneity concerns. Most importantly, the quasi-experimental design allows us to identify the causal effect of the tobacco billboard ban using appropriate causal methods. While a few causal studies exist on related products, such as e-cigarette advertising, this is, to the best of our knowledge, the first quasi-experimental study on advertising bans for tobacco.

The structure of the paper is as follows. \Cref{sec:bb:literature} reviews the literature. The individual and policy data is presented in \cref{sec:bb:data}. \Cref{sec:bb:method} outlines the identification, empirical strategy, and estimation. The main results and heterogeneous effects across subgroups are reported in \cref{sec:bb:results}, while robustness checks are discussed in \cref{sec:bb:robustChecks}. \Cref{sec:bb:conclusion} concludes.

\section{Literature}
\label{sec:bb:literature}

The literature on tobacco advertising and smoking focuses on two rationales for advertising, see for example \textcite{Saffer00, Goel06}. First, firms may advertise to increase their own market share, making current smokers switch from other brands to the firm's brand. In this case, advertising would only reshuffle existing smokers between brands while overall demand for tobacco remains unchanged. Second, firms may advertise so that people start smoking or that current smokers increase their consumption. In contrast to the previous case, advertising would increase overall demand for tobacco, resulting in an expansion of the market. We refer to this as the market expansion hypothesis. Regarding public health, advertising increasing tobacco use and subsequently reducing population health would provide a clear rationale for regulating advertising as a tobacco prevention policy.

Literature reviews report either null or positive effects of advertising on cigarette demand \parencite{Saffer00, Gallet03, Goel06}. Moreover, studies on advertising bans find null or negative effects on cigarette demand. The reviewed studies show considerable limitations regarding data and methodology. The data considered is typically a time series, aggregated at the country level, and dating prior to 2000. These constraints, along with methodological choices, may introduce endogeneity, for example, due to a lack of control for other tobacco prevention policies and structural shifts such as changes in anti-smoking sentiment. Given these limitations, the evidence on advertising remains inconclusive, as some studies support the market expansion hypothesis while others do not.

\textcite{DeCicca22} review more recent studies on advertising, using quasi-experimental approaches and micro data at the individual level. They thus tackle endogeneity concerns regarding previous studies. However, none of these studies evaluates advertising of traditional cigarettes, i.e., combustible cigarettes. They rather focus on advertising for specific products such as menthol cigarettes, e-cigarettes, and smokeless tobacco. The reviewed studies report inconclusive results as well. Some evidence is in line with the market expansion hypothesis, while other studies fail to support it.

Many pre-2000 studies evaluate the 1971 ban on television and radio cigarette advertising in the United States, typically using aggregate data \parencite{Goel06}. \textcite{Qi13} reevaluates this policy using brand-level micro data, covering over 100 cigarette brands. Compared to previous studies, the data are more fine-grained, and an innovative structural modeling approach accounts for both consumer and industry responses. On the industry side, the partial advertising ban reduced advertising efficiency, increased market concentration, and raised overall advertising spending. On the consumer side, advertising was positively associated with the share of the smoking population before the ban. After the ban, no further market expansion due to advertising is observed, indicating a shift toward competition for market share. This suggests that the partial advertising ban led to a structural change, such that the subsequently less efficient advertising no longer achieved its previously market-expanding role.

Regarding advertising bans, another relevant debate concerns the difference in effectiveness between partial and comprehensive advertising bans. Reviews of pre-2000 studies conclude that comprehensive bans covering a broad range of media types are more effective in reducing smoking rates than partial bans \parencite{Goel06, Saffer00}. One explanation for this finding could be that partial bans enable substitution into non-banned media, potentially reducing their effectiveness. Related to that argument, one may also consider that advertising has a diminishing marginal product, meaning that each additional unit of advertising spending increases smoking less than the previous one. It follows that the marginal effect of advertising on smoking may be close to zero if advertising levels are already high. Consequently, comprehensive bans restricting multiple media types are more likely to reduce smoking, in contrast to partial bans, which may leave advertising levels at a point where the marginal product remains close to zero.

\textcite{Heckman08} and \textcite{Nelson10} critique studies outside economics that claim advertising initiates youth smoking. They argue that these studies fail to identify causal effects due to multiple limitations, including specification bias, measurement error, endogenous treatments, and sample selection bias. Specification bias may arise from omitted confounding factors, such as tobacco prices, tobacco prevention policies, and family environment. Measurement error may occur because receptivity to advertising, such as ownership of tobacco-related items, is typically used as a treatment, although this measure may reflect individual preferences rather than actual exposure to advertising. Moreover, unobservable factors such as curiosity may influence both receptivity to advertising and smoking behavior, introducing endogeneity. Finally, individuals may self-select into the sample based on unobserved characteristics related to smoking behavior. 

While these challenges to causal identification are well known in economics, \textcite{Heckman08} and \textcite{Nelson10} offer valuable strategies for sound econometric estimation of the effect of advertising on smoking behavior. First, relevant confounding factors should be discussed and included, such as tobacco prices, prevention policies, and family environment. Second, robustness checks and sensitivity analyses should be conducted to assess model specification and the stability of results. Third, valid measurements of advertising and smoking behavior are required. Suitable treatments include advertising expenditures and bans, while appropriate outcome measures include smoking status and experimentation. Advertising expenditures and bans would be particularly relevant for policy, as concrete interventions can follow from them. Fourth, quasi-experimental approaches should be applied to address endogenous treatments. Fixed effects models could be beneficial for controlling unobserved heterogeneity across groups. Finally, possible selection mechanisms should be considered, as they may correlate with smoking behavior.

Our study contributes to the sparse and largely absent evidence on the causal effect of advertising expenditures and bans. It aims to provide new evidence on the causal relationship between advertising bans and smoking. Exploiting variation in advertising bans across geographic regions and over time allows us to identify their causal effect on smoking rates. The analysis is based on individual-level micro data containing more than one million observations spanning over more than two decades. State-of-the-art quasi-experimental methods, such as staggered DiD by \textcite{Callaway21} and the IFEct model by \textcite{Liu24}, are applied. Our contribution lies in conducting a thorough causal analysis based on individual-level data, improving on existing studies that suffer from considerable endogeneity concerns and rely primarily on aggregated data.

\section{Data}
\label{sec:bb:data}

\subsection{Swiss Health Survey}
\label{sec:bb:swissHealthSurvey}

We use individual data from the Swiss Health Survey, conducted by the \textcite{SGB9717}. The survey covers individuals aged 15 and older living in private households in Switzerland. It is a repeated cross-sectional survey collected every five years. The data is rich in information because not only the individuals' current smoking status but also their smoking history were collected. Where applicable, the age at smoking initiation and cessation is known. Thanks to that, we can retrospectively reconstruct each individual's smoking history from the survey interview date back to their childhood.

Smoking status at the time of the survey is known for each individual. They are asked ``Do you smoke, even if only rarely?''. If answered positively, the individual is a current smoker. Current smokers are additionally asked about the age at smoking initiation. If answered negatively, one is additionally asked about having smoked more than 6 months in one's life. If answered negatively, then one is a never smoker. If answered positively, one is a former smoker. Former smokers are then asked about their age at initiation and cessation. Consequently, never smokers, current smokers, and former smokers can be identified as well as their smoking history.

We construct a panel based on the cross-sectional data. The reconstruction of the panel is necessary because the original cross-sectional data is widely spaced in time. However, the policy change of interest occurs within relatively short time windows. For that reason, cross-sectional data alone does not provide sufficient variation in exposure to billboard bans to evaluate their effectiveness. A similar transformation from repeated cross-sectional data to a panel structure using the Swiss Health Survey was previously carried out by \textcite{Marti14} to analyze the impact of tobacco control expenditures in Switzerland. 

\begin{figure}[htbp!]
	\centering
	\includegraphics[width=0.6\textwidth]{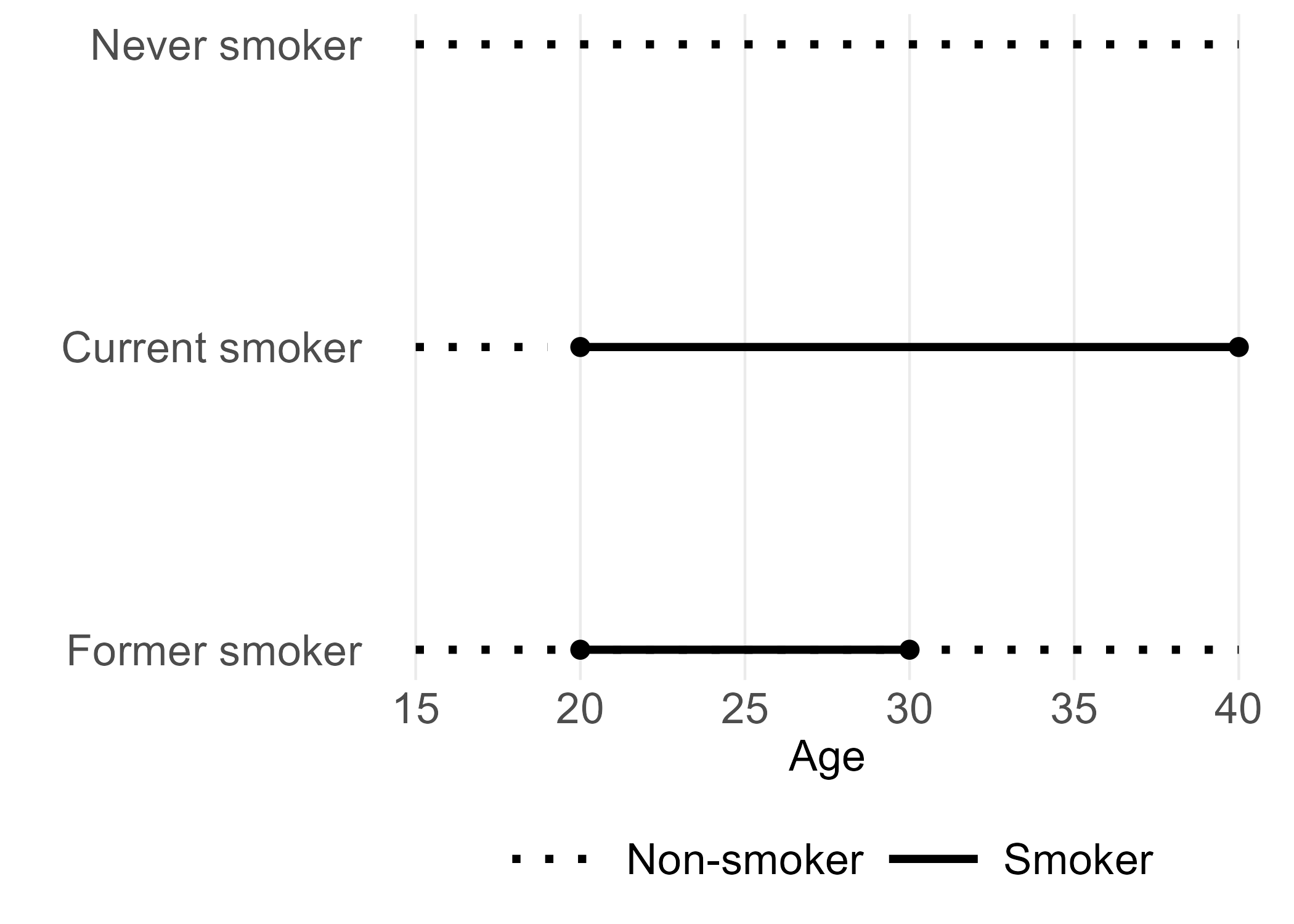}
	\caption{Smoking history of never, current, and former smokers}
	\label{fig:bb:panelConstruction}
	\caption*{\footnotesize \textit{Note: This figure illustrates the smoking history of an individual aged 40 at the time of the survey. Three scenarios are considered: (i) never smoker who never smoked in their life, (ii) current smoker who started smoking at the age of 20, and (iii) former smoker who started smoking at the age of 20 and stopped smoking at the age of 30.}}
\end{figure}

Smoking intensity in the past, i.e., the number of cigarettes smoked, is not known. For that reason, it is only known whether an individual was a smoker or non-smoker in the past. Never smokers are non-smokers throughout their lives. Current smokers are defined as smokers from the age of initiation onward (including that year) and as non-smokers before. Former smokers are defined as smokers from the age of initiation to the age of cessation (including these years), and as non-smokers before and after. \Cref{fig:bb:panelConstruction} illustrates the calculation of an individual's smoking history. 

Since 2017, surveys have recorded only an age range for smoking cessation instead of the exact age. For example, if someone reports quitting between ages 30 and 35, we know they were still smoking at age 30 because it marks the beginning of the range. However, their smoking status from age 31 to 35 is uncertain, as they could have quit at any point within that range. These years are therefore coded as missing, since the exact smoking status cannot be determined.

The panel is constructed using survey data from the years 1997, 2002, 2007, 2012, and 2017. The most recent 2022 survey is omitted to avoid potential limitations due to the COVID-19 pandemic. The five selected survey waves contain a total of 95,200 individual observations. Of these, 2.8\% are excluded due to missing or ambiguous smoking status, where individuals could not be identified as never smokers, current smokers, or former smokers. Each individual's smoking history is retrospectively reconstructed going back to age 15, or the year 1993 at the latest. The resulting data is an unbalanced panel with 1,268,266 observations from 1993 to 2017. As we know the complete smoking history of the individuals, the smoking status in each year before their survey interview can be recalculated. The panel is unbalanced because each survey year samples a different set of individuals.

Constructing the retrospective panel results in the loss of time-varying socio-demographic information. However, this limitation is offset by a substantial increase in the number of observations and annual data points. Yearly information is particularly important because the policy of interest, billboard bans, was introduced in different regions within a short time frame. Apart from smoking status, only age can be directly reconstructed. Assuming gender does not change over time, an individual's gender at the survey date is treated as a time-invariant covariate. The resulting panel therefore includes yearly information on individuals’ smoking status, gender, and age. 

\Cref{tab:bb:comparisonComposition} compares the sample composition of the original cross-sectional data and the reconstructed panel. The gender distribution remains largely unchanged. However, because age is calculated retrospectively, the panel includes a higher share of younger individuals. \Cref{tab:bb:comparisonSmoking} compares smoking rates in the total sample and across subgroups before and after panel reconstruction. In the total sample, differences are moderate and show no consistent direction. The smoking rates are sometimes higher and sometimes lower in the panel. These discrepancies are similar across genders. However, by age group, discrepancies tend to be more positive among individuals aged 25--64, and more negative among younger and older groups. This suggests that, compared to the original data, smoking rates are overestimated for the middle age groups and underestimated for the youngest and oldest in the panel. 

\FloatBarrier

\subsection{Tobacco Prevention Policies}
\label{sec:bb:tobaccoPreventionPolicies}

In Switzerland, tobacco prevention policies and programs are implemented at multiple levels of government. Switzerland consists of 26 states, called cantons, which can apply their own policies within their region. Major tobacco prevention policies are implemented at the national and cantonal level. National policies affect the entire population of Switzerland, while cantonal policies only affect specific geographic regions of Switzerland. Besides the government, other organizations promoting health also play a major role in tobacco prevention. There is no systematic collection of tobacco prevention policies and programs in Switzerland. It is thus challenging to capture all information on tobacco prevention. 

Several policies are enacted at the national level, most notably tobacco taxes, smoking bans, health warnings, and bans on advertising on television and radio \parencite{BAG25swiss}. Prior to 2024, however, national legislation played a relatively minor role, and tobacco prevention efforts were largely driven by the cantons. In 2024, many important tobacco and nicotine prevention policies that had previously existed at the cantonal level were adopted at the national level, including the tobacco billboard bans, sales bans for minors, and the extension of tobacco regulations to e-cigarettes \parencite{BAG25swiss}. Despite their importance, these national laws have limited relevance for the empirical strategy of this DiD study, as they do not vary across cantons.

We focus on cantonal tobacco prevention policies for two reasons. First, policy data is available, in contrast to information on programs and interventions. Second, our policy of interest in this analysis, i.e., the billboard bans, is a cantonal policy. Hence, other cantonal tobacco prevention policies are important confounding factors. The \textcite{BAG25cantonal} provides information on the most relevant cantonal tobacco prevention policies, including advertising restrictions, sales bans, and smoking bans. These policies were introduced at different times across cantons, while some cantons have not implemented them. The cantonal policies often follow a similar legal wording and thus a similar set of restrictions. But cantons vary in the intensity of the policy, implementing only partial or more comprehensive restrictions. 

The policy data from the \textcite{BAG25cantonal} has an important constraint. It only records the most recent amendment date. For example, if a canton introduced a sales ban for minors under the age of 16 in 2010 and extended it to those under 18 in 2020, the source would list 2020 as the implementation year. We therefore reviewed all relevant cantonal laws and regulations to reconstruct the complete policy history. The law collection of each of the 26 cantons had to be consulted. As a result of this effort, we compiled a policy dataset that captures the full policy history. It covers three major cantonal tobacco prevention policies, namely tobacco billboard, sales, and smoking bans. 

\Cref{tab:bb:cantonalPolicies} lists the implementation dates of cantonal billboard, sales, and smoking bans, based on our unique policy dataset. The date for the billboard ban refers to when such a ban was first introduced either in public areas or in areas visible from public spaces. The sales ban date corresponds to the earliest date on which the sale of tobacco to minors under the age of 16 or 18 was prohibited. The smoking ban date indicates when any restriction on smoking in public areas was first adopted.

\begin{figure}[htbp!]
	\centering
	\includegraphics[width=0.6\textwidth]{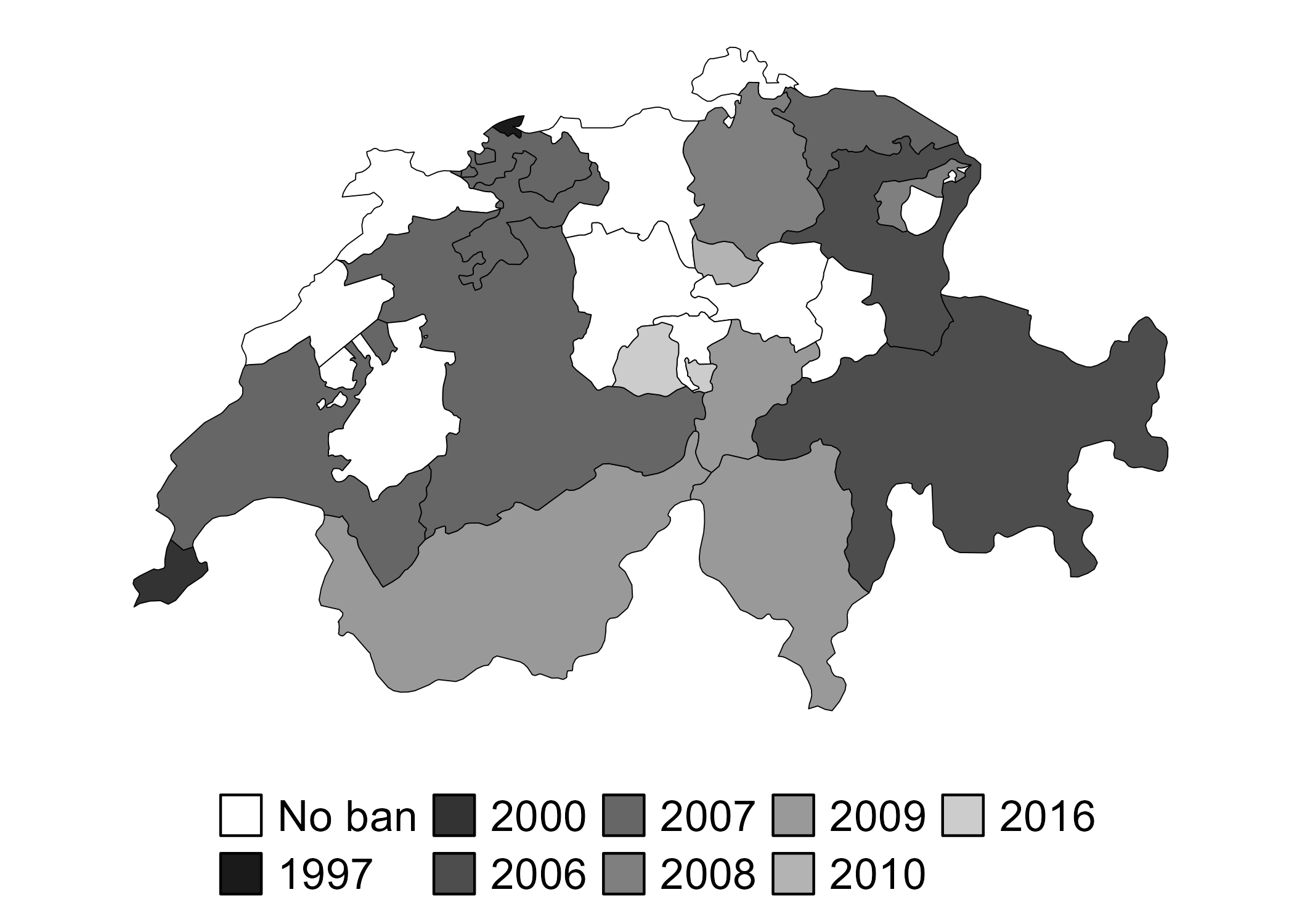}
	\caption{Introduction of cantonal tobacco billboard bans}
	\label{fig:bb:billboardBans}
	\caption*{\footnotesize \textit{Note: The plot depicts the introduction of cantonal tobacco billboard bans between 1993 and 2017, indicating the year of adoption for each canton.}}
\end{figure}

\FloatBarrier

\Cref{fig:bb:billboardBans} provides a geographic representation of the cantons with a tobacco billboard ban as of the end of 2017. Two early adopters introduced such bans in 1997 in the canton of Basel-Stadt and in 2000 in the canton of Geneva. The introduction in Geneva was highly contested \parencite{Kuenzler18}. In contrast to the ban in Basel-Stadt, the Geneva regulation extended to billboards visible from public areas, thereby including those located on private property. A federal court decision in 2002 confirmed that the Geneva billboard ban was lawful and that cantons were permitted to enact this type of preventive policy. The example was followed by many other cantons, particularly between 2005 and 2010. By 2010, around three-quarters of the Swiss population was already covered by a billboard ban \parencite{BAG15}. In 2017, 16 out of 26 cantons adopted a billboard ban. Finally, the graphical depictions highlight that policy introduction varies across time and regions, which is an ideal quasi-experimental setting.

\section{Method}
\label{sec:bb:method}

\subsection{Identification Strategy}
\label{sec:bb:idStrategy}

The identification strategy builds on the potential outcomes framework \parencite{Neyman23, Rubin74}. The conventional DiD approach considers a setting with two time periods and two groups, as discussed in \textcite{Abadie05, Lechner11}. In contrast, our analysis involves multiple time periods and variation in treatment timing because different groups adopt the policy at different points in time. This staggered adoption introduces the possibility of heterogeneous treatment effects across these groups. Accordingly, the identification strategy must account for such heterogeneity as discussed in the following. Further implications for estimation are presented in \cref{sec:bb:estInf}.

The main analysis is based on the staggered DiD approach by \textcite{Callaway21}, which is well-suited for settings with multiple time periods and variation in treatment timing. We closely follow their identification strategy and notation. The core identification assumptions remain similar to those of the conventional DiD framework. The approach aims to identify the average treatment effect on the treated (ATET) for each treatment group and time period combination. The group-time-specific ATET, denoted as $ATET(g, t)$, refers to the ATET for the group first treated in period $g$, evaluated in time period $t$. The causal parameter of interest is thus defined at a more granular level, compared to the conventional two-way fixed effects estimator, which averages over treatment groups and time periods. Staggered DiD accounts for treatment effect heterogeneity precisely because it estimates these group-time-specific ATETs separately in a first step. In a second step, these group-time-specific ATETs can then be aggregated to obtain more informative overall estimates, as discussed in \cref{sec:bb:estInf}.

The identification assumptions required to identify the group-time-specific ATET are developed in the remainder. We consider a setting with multiple individuals $\mathcal{I}$, indexed by $i \in \mathcal{I}$, and multiple time periods $\mathcal{T}$, indexed by $t = 1, \dots, \mathcal{T}$. Let $D_{i,t} \in \{0,1\}$ denote a binary treatment indicator, where $D_{i,t} = 1$ if individual $i$ is treated in period $t$, and $D_{i,t} = 0$ otherwise. We assume that the treatment is staggered and absorbing. This implies that (i) no individual is treated in the initial period $t = 1$, (ii) individuals may begin treatment in different periods, and (iii) once treated, individuals remain treated in all subsequent periods.

Treated individuals can be grouped according to the period in which they first receive treatment. Let $G_g = \mathbf{1}\{G = g\}$ denote a binary indicator equal to one if a unit is first treated in period $g$. Let $Y_{i,t}(0)$ denote the potential outcome of individual $i$ at time $t$ if never treated, and let $Y_{i,t}(g)$ denote the potential outcome at time $t$ if first treated in period $g$, for each $g \in \{2, \dots, \mathcal{T}\}$. Under the observational rule, the observed outcome can be expressed as $Y_{i,t} = Y_{i,t}(0) + \sum_{g=2}^{\mathcal{T}} \left( Y_{i,t}(g) - Y_{i,t}(0) \right) \cdot G_{i,g}$. For never-treated individuals, observed outcomes equal the untreated potential outcomes in all periods. For individuals first treated in period $g$, observed outcomes equal the untreated potential outcomes plus the treatment effect from period $g$ onward.

We aim to identify the group-time ATET, defined as:
\begin{align}
	ATET(g,t)=\mathbb{E}[Y_t(g) - Y_t(0) | G_g = 1]. \label{eq:bb:gtATET}
\end{align}
This parameter captures the ATET for individuals first treated in period $g$, evaluated in time period $t$. The group-time-specific ATET in \cref{eq:bb:gtATET} is identified under the following assumptions. 

First, we impose a no anticipation assumption, which states that treatment assignment does not affect potential outcomes in pretreatment periods. Formally,
\begin{align}
	\mathbb{E}[Y_t(g) | X, G_g = 1] = \mathbb{E}[Y_t(0) | X, G_g = 1], \forall t < g, 
\end{align}
where $X$ denotes observed covariates used to control for confounding.

Second, we assume conditional parallel trends based on the not-yet-treated groups. Individuals who have not yet received treatment by period $t$, but who will be treated in a later period $g > t$, are considered part of the not-yet-treated group. While one could alternatively base the parallel trends assumption on the never-treated group, the not-yet-treated group appears to be more appropriate as a control. First, the never-treated group is relatively small, as most units eventually receive treatment. Second, not-yet-treated individuals may be more comparable in terms of socio-demographic characteristics, as they are from regions that ultimately adopt the billboard ban. The conditional common trends assumption requires that, in the absence of treatment, the trends in mean potential outcomes are identical among the treatment and control groups, conditional on covariates $X$. This assumption must hold separately within each treatment group $G_g$, i.e., the group of units first treated in period $g$. Formally,
\begin{align}
	\mathbb{E}[Y_t(0) - Y_{t-1}(0) | X, G_g = 1] = \mathbb{E}[Y_t(0) - Y_{t-1}(0) | X, D_t = 0, G_g \neq 0], \forall t \geq g. \label{eq:bb:commonTrend}
\end{align}
Finally, two additional assumptions are required due to the inclusion of covariates. First, we impose a common support condition to ensure that treatment assignment is not deterministic. Specifically, for each group treated in period $g$ and for each period $t \in \{2, \dots, \mathcal{T}\}$, the treatment propensity score conditional on covariates $X$, denoted by $p_{g,t}(X) = Pr(G_g=1 | X, G_g + (1-D_t) (1-G_g) = 1)$, must satisfy $p_{g,t}(X) < 1$. This ensures overlap in covariate distributions between treatment and control groups. Second, we assume that treatment does not affect the covariates $X$, also referred to as the exogeneity of confounders assumption.

We additionally consider the IFEct estimator by \textcite{Liu24}, which is a latent factor model. They build on earlier factor models developed by \textcite{Gobillon16} and \textcite{Xu17}. This approach estimates latent unit-specific loadings and time-specific factors based on treatment and control units. Based on this latent factor model, potential outcomes under nontreatment are then imputed for treated units in posttreatment periods to estimate treatment effects. Similar to the staggered DiD approach, this estimator is robust to heterogeneous treatment effects. In contrast, IFEct estimators allow for violation of the common trends assumption. This feature is the main motivation for introducing this second estimator, as discussed in \cref{sec:bb:empStrategy}. The inclusion of IFEct estimators in the DiD textbook by \textcite{DeChaisemartin23} shows that they are increasingly relevant for applied economics research and will likely be used more in the future.

Building on the previously defined notions, we consider individuals $i \in \mathcal{I}$, time periods $t \in \mathcal{T}$, treatment status $D_{i,t} \in \{0,1\}$, and potential outcomes $Y_{i,t}(1)$ and $Y_{i,t}(0)$. Let $X_{i,t}$ denote observed covariates and $U_{i,t}$ denote unobserved attributes for unit $i$ at time period $t$. Based on these components, the potential outcome under nontreatment for individual $i$ at time period $t$ is modeled as:
\begin{align}
	Y_{i,t}(0) = f(X_{i,t}) + h(U_{i,t}) + \varepsilon_{i,t}, \label{eq:bb:imputedOutcome}
\end{align}
where $f(\cdot)$ and $h(\cdot)$ are functions capturing the effects of observed and unobserved components, respectively, and $\varepsilon_{i,t}$ is an idiosyncratic error term. The unobserved component is based on latent unit-specific loadings and time-specific factors, capturing unobserved time-varying heterogeneity and thereby relaxing the common trends assumption.

The causal parameter of interest is the ATET for individual $i$ at time period $t$:
\begin{align}
	ATET(i, t) = \mathbb{E}[Y_{it}(1) - Y_{it}(0) | D_{it} = 1], \forall t > T_0, \label{eq:bb:ifectATET}
\end{align}
where $T_0$ denotes the policy introduction period. The ATET is defined only for treated units in posttreatment periods, excluding always- and never-treated units. 

The causal effect is identified under the following three main assumptions. First, we make a functional form assumption as stated in \cref{eq:bb:imputedOutcome}. Accordingly, the potential outcome under nontreatment consists of three additive components, namely observed covariates, unobserved attributes, and idiosyncratic error. Similar to the assumptions of the staggered DiD approach, this implies no anticipation of the treatment. Second, strict exogeneity requires that the treatment $D_{it}$ is independent of the idiosyncratic error term, conditional on the observed $X_{it}$ and unobserved $U_{it}$ components. Third, the unobserved component $h(U_{it})$ is assumed to be approximable by a low-rank matrix, meaning that the number of latent factors and unit-specific loadings is sufficiently small relative to the sample size. In contrast to the staggered DiD approach, common trends are not required for identification because the factor model allows for more flexible confounding.

\subsection{Empirical Strategy}
\label{sec:bb:empStrategy}

The staggered introduction of billboard bans by the cantons, i.e., geographic subregions of Switzerland, resulted in variation across regions and over time. Exploiting this quasi-experimental setting, we estimate the effect of billboard bans on smoking. Prior to the widespread adoption of these bans between 2005 and 2010, the vast majority of tobacco advertising expenditure in Switzerland was allocated to billboards \parencite{BAG15}. Until 2005, the industry consistently spent around 45 million Swiss francs annually on billboard advertising in most years. Spending declined sharply thereafter and stabilized at approximately 5 million Swiss francs per year from 2010 onward. Overall advertising expenditure fell by more than half, from 49.9 to 16.8 million Swiss francs annually between 2005 and 2010. Substitution to other media was limited, with newspaper advertising being the only notable shift. This suggests that billboards were not only important but also an effective media type that proved difficult to replace. Given billboards' central role and their limited substitution, the billboard bans are expected to have a strong impact and remain one of the most relevant advertising restrictions to study in the Swiss context.

The identification of the causal effect of billboard bans is challenging due to Switzerland’s complex tobacco policy landscape. At the national level, tobacco advertising on radio and television was banned as early as the 1960s \parencite{BAG15}. In 1995, advertising targeted at minors and aimed at persuading them to smoke was also prohibited. In addition to these restrictions, Switzerland introduced multiple prevention policies. Most importantly, indoor smoking bans, tobacco taxes, and health warnings were implemented at the national level \parencite{BAG25swiss}. Although adopted much later than the evaluated cantonal billboard bans, the new tobacco product law introduced a wide range of additional national policies in 2024. Because these policies affect the entire Swiss population, they pose a limited threat to the identification of the cantonal billboard bans, which were only adopted in specific subregions of Switzerland.

More concerning for identification are the subregional policies at the cantonal level. In fact, cantons are allowed to introduce stricter advertising bans, for example by extending them to billboards, cinemas, or events \parencite{BAG25cantonal}. The billboard ban was the most widely adopted restriction, while some cantons additionally extended these bans to cinema advertising and sponsorship. In addition, cantons differ in their tobacco sales bans for minors and in the implementation of indoor smoking bans. Because the billboard bans assessed in this study were introduced at the same institutional level, they may be confounded by these other cantonal tobacco prevention policies.

We control for cantonal sales and smoking bans to account for confounding. Because no comprehensive collection of tobacco prevention policies and programs exists, we focus on cantonal policies, which account for most of the variation across cantons. Sales and smoking bans are both widely adopted and can be clearly attributed to individual cantons, making them relevant and feasible control variables. In addition, individuals' gender and age are included to account for potential differences in trends of population composition between treatment and control groups. The individual-level data also presents limitations due to the retrospective panel reconstruction, which results in the loss of time-varying information. In summary, we control for relevant confounders, namely cantonal sales and smoking bans, as well as individuals’ gender and age, but the ability to control for confounding remains limited due to the stated data constraints. 

Limited control may lead to violations of the conditional common trends assumption. As discussed in \cref{sec:bb:resultsMain}, this identification assumption may not hold for the results based on the staggered DiD approach. Some unobserved confounding factors may remain that are not captured by the included controls. This motivates the use of the IFEct estimator, which accounts for latent confounding by modeling unit- and time-specific trends. Unlike staggered DiD, this estimator allows for violations of the common trends assumption. Given the data constraints and the likelihood of remaining confounding, the IFEct estimator may be more suitable for this application, as also indicated by the results in \cref{sec:bb:resultsMain}. However, both estimators are considered complementary, since the validity of identification assumptions cannot be definitively proven.

Besides confounding, there are remaining challenges to identification due to the panel reconstruction and other factors. First, due to the panel reconstruction, time-varying information on the individuals is lost. Hence, residency changes of the individuals before the survey date are unknown. In \cref{sec:bb:robustChecks}, we consider individuals' smoking histories over different periods of time, indicating that the bias due to residency changes is negligible. Second, mobility across cantons may lead to partial billboard ban exposure of the control group and vice versa. Assuming a smoking reduction due to billboard bans, this would suggest that we would underestimate the treatment effect, i.e., estimate a lower bound. In \cref{sec:bb:robustChecks}, spillover effects are investigated and found to attenuate the estimated effects. Third, reverse causality may lead cantons with high smoking rates to introduce stricter advertising restrictions, including billboard bans. This would imply a positive bias in the treatment effect and suggest again that a lower bound would be estimated. Fourth, discussion preceding the billboard ban may already affect smoking, which would violate the no anticipation assumption. But placebo tests in pretreatment periods suggest that such anticipation effects are not present, see \cref{sec:bb:resultsHet}. In \cref{sec:bb:robustChecks}, the sensitivity of our results to alternative model specifications is tested, addressing potential limitations. 

Finally, retailers may respond to billboard bans by reducing cigarette prices in affected cantons in an attempt to retain customers and maintain sales. The retail price of a pack of 20 cigarettes in Switzerland was 8.50 Swiss francs in 2017 \parencite{BAZG26}, with taxes accounting for the majority of the price. Although a single reference price is typically cited, tobacco prevention advocacy organizations note that discounting practices can result in considerably lower effective prices \parencite{ATSchweiz26}. However, the large tax component, together with manufacturer and distribution margins, likely leaves limited room for retailer profits, making a sustained price response unlikely. Even if such a response were observed, it would bias our estimates toward zero, as lower cigarette prices in affected cantons would partially offset the effect of billboard bans on smoking. Overall, these remaining limitations are likely to attenuate the estimated treatment effects, suggesting conservative estimates that represent a lower bound.

\subsection{Estimation and Inference}
\label{sec:bb:estInf}

Estimation is based on the staggered DiD estimator proposed by \textcite{Callaway21}, which is well-suited for settings with multiple time periods and variation in treatment timing. This estimator is preferred over two-way fixed effects approaches, which may yield biased estimates in the presence of treatment effect heterogeneity, as discussed in \textcite{Borusyak24, DeChaisemartin20, GoodmanBacon21, Sun21}. The bias arises because already-treated units may serve as inappropriate controls for later-treated units. To address this issue, staggered DiD estimators, such as the one suggested by \textcite{Callaway21}, have been developed that are robust to heterogeneous treatment effects.

Their estimator additionally allows for absorbing treatments and the inclusion of covariates, which is required in our empirical setting. Covariates must be included because common trends may not hold unconditionally. Moreover, once a treatment group is exposed to the billboard ban, it remains treated in all subsequent periods, constituting an absorbing treatment state. By combining inverse probability weighting with an outcome regression model, their estimator is not only robust to treatment effect heterogeneity but also doubly robust. As a result, it yields consistent estimates if either the treatment propensity score model or the outcome model is correctly specified.

To avoid incorrect comparisons, the estimator first calculates all group-time-specific ATETs, $ATET(g, t)$, as defined in \cref{eq:bb:gtATET}. These correspond to the ATET for units first treated in period $g$, evaluated at time $t \geq g$. In a second step, these group-time-specific ATETs are aggregated across groups and time periods to obtain meaningful overall effects. We follow an event-study design, aggregating the group-time-specific ATETs relative to the time of policy introduction. This allows us to investigate dynamic treatment effects by exposure length, i.e., event time $e = t - g$. The average group-time-specific ATETs, conditional on time elapsed since policy introduction $e$, are given by:
\begin{align}
	\widehat{ATET}(e) = \sum_{g:g+e < \mathcal{T}} \Pr(G_g=1| G+e \leq \mathcal{T}) \widehat{ATET}(g, g+e),
\end{align}
with $e\in \{0,1,\ldots ,\mathcal{T}-1\}$. The event-study estimates are the group-time-specific ATETs with exposure length $e$ summed up across all treatment groups $g$ and weighted according to the share of treated units still observed after exposure length $e$, $\Pr(G_g=1| G+e \leq \mathcal{T})$. We refer to this as the time-specific, or year-specific, effect, as it captures the treatment effect $e$ periods, or years, after policy introduction. Finally, they can be averaged over multiple exposure lengths, allowing us to report average time-specific effects from the time of policy introduction up to event time $e \in \{0, 1, \ldots, \mathcal{T}-1\}$.

The IFEct estimator allows for the construction of event-study estimates in a similar fashion. Here, the time-specific ATETs are aggregated based on individual treatment effects. For estimation, let $f_t$ denote the time-specific factors, and $\lambda_i$ the unit-specific loadings. In a first step, the time-specific factors are learned based on all control units. In a second step, the unit-specific loadings are estimated among the treated units in the pretreatment periods. Finally, potential outcomes under nontreatment are estimated for unit $i$ at time $t$:
\begin{align}
	Y_{it}(0) = X_{it}'\beta + \alpha_i + \xi_t + \lambda_i' f_t + \varepsilon_{it}, 
\end{align}
where $X_{it}$ are covariates, $\alpha_i$ and $\xi_t$ are unit and time fixed effects, $\lambda_i' f_t$ are interacted time-specific factors and unit-specific loadings, and $\varepsilon_{it}$ is the idiosyncratic error term. The counterfactual outcome is imputed based on covariates, fixed effects, and, most importantly, the interacted latent factors and loadings. Note that the interaction allows for unit-specific trends, which ultimately relaxes the common trends assumption.

Individual treatment effects are defined for treated units in the posttreatment period, excluding always- and never-treated units. For each treated unit $i$, the potential outcome under nontreatment is imputed. Based on the observed outcome of the treated unit and the imputed outcome under nontreatment, the individual treatment effect is calculated as $\hat{\delta}_{it} = Y_{it} - \hat{Y}_{it}(0)$. The individual treatment effects are then aggregated to time-specific ATETs relative to the time since treatment introduction $e$:
\begin{align}
	\widehat{ATET}(e) = \frac{1}{N_e} \sum_{i \in \mathcal{I}_e} \hat{\delta}_{it} = \frac{1}{N_e} \sum_{i \in \mathcal{I}_e} Y_{it} - \hat{Y}_{it}(0),
\end{align}
where $\mathcal{I}_e$ is the set of treated units observed $e$ periods after their treatment began, and $N_e = |\mathcal{I}_e|$ is the number of such units.

Event-study estimates reporting the year-specific ATETs of the billboard ban on smoking constitute the main results. The main results are based on two estimators, the staggered DiD approach and the IFEct estimator. For both approaches, the whole sample and all available covariates are used, including socio-demographic characteristics, i.e., gender and age, as well as tobacco prevention policy variables, i.e., tobacco sales and smoking bans. By construction, the IFEct estimator does not consider time-invariant covariates, as it already accounts for unit fixed effects. Standard errors are clustered at the individual level in both cases.

The staggered DiD estimation is conducted in Stata using the \texttt{csdid} package by \textcite{Rios23}. The improved doubly robust estimator with inverse probability tilting is applied, setting the method to \texttt{drimp}. Not-yet-treated units are selected as the control group. A robustness check using never-treated units as the control group yields similar results and is provided in \cref{sec:bb:robustChecks}. The IFEct estimation is based on the Stata package \texttt{fect} by \textcite{Liu24}, setting the method to \texttt{ife} and imposing two-way fixed effects. Nonparametric standard errors are obtained through 200 bootstrap replications. Cross-validation is used to select the optimal number of latent time-specific factors.

\section{Results}
\label{sec:bb:results}

\subsection{Effect of Tobacco Billboard Bans}
\label{sec:bb:resultsMain}

\begin{figure}[htbp!]
	\centering
	\includegraphics[width=0.6\textwidth]{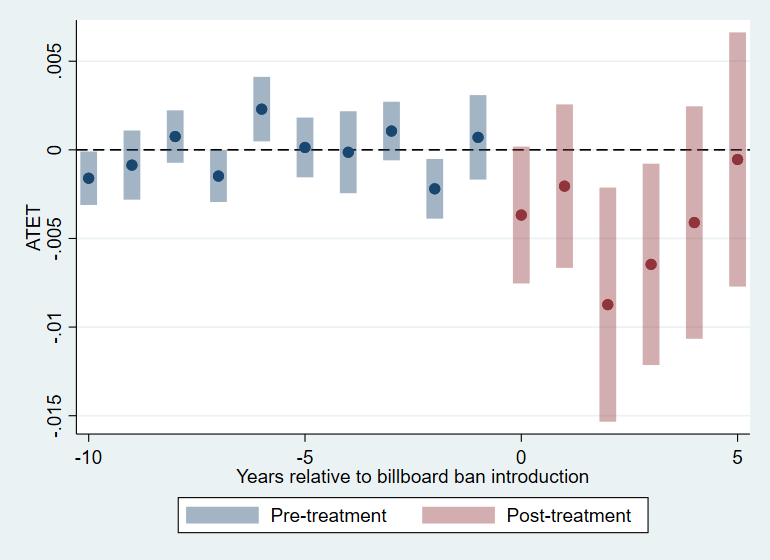}
	\caption{Year-specific effects of tobacco billboard bans on the smoking rate}
	\label{fig:bb:mainResNotYet}
	\caption*{\footnotesize \textit{Note: This figure presents year-specific ATETs of tobacco billboard bans on the smoking rate, estimated using the staggered DiD approach by \textcite{Callaway21}. Event time zero corresponds to the year of billboard ban implementation. The specification includes all covariates and cluster-robust standard errors. The point estimates are shown as dots, with 95\% confidence intervals represented by error bars.}}
\end{figure}

\Cref{fig:bb:mainResNotYet} presents the event-study estimates of the effect of tobacco billboard bans on smoking rates, using the staggered DiD approach proposed by \textcite{Callaway21}. We estimate year-specific ATETs for up to ten years before and five years after the introduction of the billboard ban. The broad time window allows us to investigate dynamic treatment effects across varying treatment exposure lengths and to assess identification assumptions through placebo tests. These estimates are based on the main specification stated in \cref{sec:bb:estInf}, which includes all covariates and uses cluster-robust standard errors at the individual level.

The year-specific ATET in the year of the billboard ban introduction is -0.4 pp (p $=$ 0.06). The year-specific ATETs after two years (-0.9 pp, p $=$ 0.01) and three years (-0.6 pp, p $=$ 0.03) are higher in magnitude and statistically significant at the 5\% level. Also, the ATETs one year (-0.2 pp, p $=$ 0.38), four years (-0.4 pp, p $=$ 0.22), and five years (-0.1 pp, p $=$ 0.88) after policy introduction suggest a reduction in the smoking rate, but are not significant at the 5\% level.

Column (4) of \cref{tab:bb:mainRes} reports year-specific ATETs averaged over one, three, and five years following policy implementation, including the year-specific ATET of the implementation year. The average posttreatment effect over 0--1 years after billboard ban introduction is -0.3 pp (p $=$ 0.11), -0.5 pp (p $=$ 0.02) over 0--3 years after adoption, and -0.4 pp (p $=$ 0.08) over 0--5 years after implementation. Overall, the results suggest a reduction in the smoking rate due to the billboard bans, with an average decrease of 0.4 pp over five years following their adoption.

\Cref{fig:bb:mainResNotYet} additionally reports the year-specific ATETs for the pretreatment periods. While the ATETs in the posttreatment periods consistently use the year immediately preceding the introduction of the billboard ban as the reference period, this is not the case for the pretreatment ATETs. Instead, they compare outcomes between one and two years before adoption, two and three years before adoption, and so forth. This approach is referred to as short-gaps estimation. For long-gaps estimation of the ATETs in the pretreatment periods, see \cref{sec:bb:robustChecks}.

We assess the conditional common trends assumption with placebo tests using ten years of pretreatment periods. Column (4) of \cref{tab:bb:mainRes} reports the year-specific ATET immediately preceding policy implementation, as well as the ATETs averaged over five and ten years before policy introduction. The estimate for the year directly before implementation is particularly important, as all posttreatment effects are interpreted relative to this baseline. The placebo effect immediately before adoption is with a p-value of 0.56 not significant at the 5\% level. Also, the average placebo effects over 5 years (p $=$ 0.84) and over 10 years (p $=$ 0.61) are not significant. 

The placebo tests based on the event-study estimates generally support the validity of the conditional common trends assumption. However, the placebo tests do not hold in some pretreatment periods. Moreover, the placebo tests do not hold within every treatment group, i.e., groups that adopted the billboard ban at different time periods. The latter would be required for the conditional common trends assumption to hold, as defined in \cref{eq:bb:commonTrend}. This indicates that the validity of the assumption is limited. Due to this limitation, we also employ an alternative estimator that allows for violations of the common trends assumption.

\begin{figure}[htbp!]
	\centering
	\includegraphics[width=0.6\textwidth]{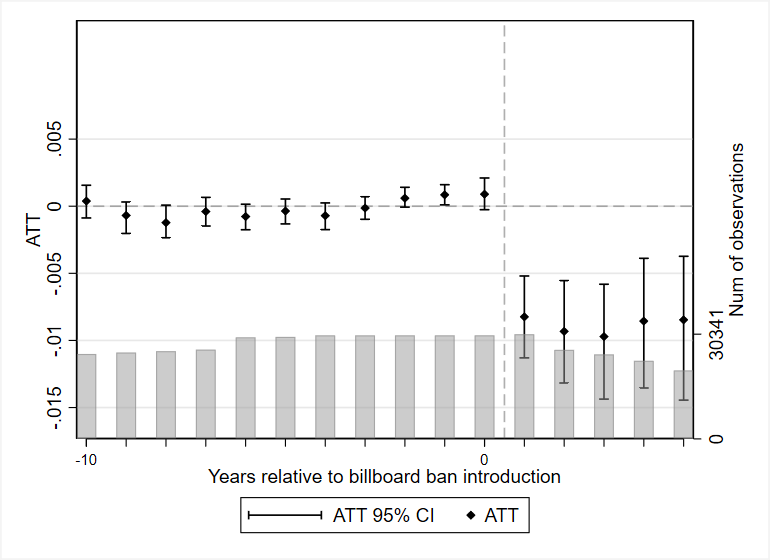}
	\caption{Year-specific effects of tobacco billboard bans on the smoking rate}
	\label{fig:bb:ifeMainRes}
	\caption*{\footnotesize \textit{Note: This figure presents year-specific ATETs of tobacco billboard bans on the smoking rate, estimated using the IFEct estimator by \textcite{Liu24}. Event time zero corresponds to the year of billboard ban implementation. The specification includes all covariates and cluster-robust standard errors. The point estimates are shown as dots, with 95\% confidence intervals represented by error bars. The gray bars at the bottom show the number of included treated observations for the corresponding point estimate above.}}
\end{figure}

\Cref{fig:bb:ifeMainRes} shows the event-study estimates based on the IFEct estimator by \textcite{Liu24}. Year-specific ATETs are estimated up to ten years before and five years after its implementation. The number of treated observations is also shown. Note that the number of observations varies considerably due to the panel being unbalanced. The cross-validation procedure selected one factor, indicating the presence of unobserved time-varying confounding, which can be mitigated through the latent factor model. The main specification was kept, except for time-constant covariates, which are not compatible with this approach.

Year-specific ATETs consistently show a statistically significant reduction in smoking rates following the billboard ban. In the year of implementation, the ATET is -0.8 pp (p $<$ 0.01), indicating an immediate reduction in the smoking rate. The effect remains negative and statistically significant in the subsequent years: one year (-0.9 pp, p $<$ 0.01), two years (-1.0 pp, p $<$ 0.01), three years (-0.9 pp, p $<$ 0.01), and four years (-0.8 pp, p $<$ 0.01) after billboard ban introduction. These results would suggest that the billboard ban led to both immediate and sustained reductions in smoking rates.

The year-specific ATETs across the pretreatment periods are close to zero and not significant at the 5\% level. This indicates that the latent factor model effectively accounts for unobserved time-varying confounding. The number of treated observations remains stable, particularly over the first seven pretreatment years and in the first posttreatment year. Afterwards, the number of treated observations decreases from 30,341 in the year of the billboard ban introduction to 19,856 four years after its implementation. This change in sample composition may limit the validity of later year-specific ATETs in posttreatment years.

The effectiveness of billboard bans was assessed based on the staggered DiD and IFEct estimators. There are trade-offs to consider, as each approach relies on a different set of assumptions. While the DiD approach requires common trends, the IFEct estimator relaxes this assumption. Relaxing common trends may appear advantageous, especially given probable unobserved confounding, e.g., unobserved tobacco prevention policies and programs. However, the IFEct estimates rely on multiple pretreatment periods due to the latent factor model, whereas the DiD estimates are relative to the first pretreatment period only. Relying on multiple pretreatment periods may be disadvantageous because population composition changes over time. As the actual data generating process is unknown, these assumptions cannot be definitively proven. Both approaches rely on different identifying assumptions that are not nested. For that reason, their estimates should be understood as complementary. We interpret consistent results in effect direction and size as an indication of an actual policy effect.

Concluding, the IFEct estimates suggest an immediate reduction of around 0.9 pp that persists over time, while the staggered DiD estimates are less pronounced, with an average reduction of 0.4 pp over five years following the billboard ban introduction. Overall, both approaches yield consistent results and therefore indicate a reduction in smoking rates due to the billboard ban.

\FloatBarrier

\subsection{Placebo Analysis}
\label{sec:bb:placeboAnalysis}

The validity of the DiD design is tested using placebo event-study estimates. Placebo treatments are constructed by shifting the actual treatment 5, 6, and 7 years before its introduction, while keeping all other specifications unchanged. Because several posttreatment periods are evaluated, the treatment must be shifted back by multiple years. This ensures that the placebo event-study estimates are not contaminated by actual treatment effects. Since no actual treatment occurred during these placebo posttreatment periods, the corresponding placebo event-study estimates are expected to be indistinguishable from zero.

\Cref{fig:bb:mainResPlacebo} shows the placebo event-study estimates for both the staggered DiD and IFEct estimators. For the staggered DiD estimator, the year-specific placebo estimates are generally small in magnitude, typically not exceeding $\pm 0.5$ percentage points, and their confidence intervals include zero. \Cref{fig:bb:mainResPlacebo} (c) is an exception, where several year-specific placebo estimates are positive and statistically significant at the 5\% level. For the IFEct estimator, the year-specific placebo estimates are again generally small in magnitude, none exceeding $\pm 0.5$ percentage points, and their confidence intervals include zero. \Cref{fig:bb:mainResPlacebo} (b) is an exception, where two year-specific placebo estimates are positive and statistically significant at the 5\% level. However, shifting treatment dates earlier reduces the number of observations available to estimate the latent factor model. The fluctuating pretreatment estimates indicate that the model fit is not sufficient. Overall, the placebo analysis does not reveal any systematic placebo effects that would invalidate the DiD design.

\FloatBarrier

\subsection{Effect Heterogeneity by Gender and Age}
\label{sec:bb:resultsHet}

Responses to tobacco advertising may vary across subgroups and appear to be gendered and age-dependent \parencite{Heckman08, Nelson10}. While our analysis focuses on heterogeneous effects by gender and age, such effects may also arise due to socio-economic status, family environment, and other factors. \Cref{fig:bb:hetRes} presents event-study estimates of the effect of tobacco billboard bans on the smoking rate, stratified by gender (men and women) and age (15--24, 25--44, 45--64, and 65+ years old), using the staggered DiD estimator by \textcite{Callaway21}. As with the previous results, \cref{tab:bb:hetRes} summarizes average year-specific ATETs over one, three, and five years after billboard ban implementation across these subgroups. 

We find effect heterogeneity both with respect to gender and age. The average year-specific ATETs for women are -0.8 pp (p $<$ 0.01) over one year, -1.0 pp (p $<$ 0.01) over three years, and -0.9 pp (p $<$ 0.01) over five years after billboard ban implementation. Regarding age groups, the effect among 25--44-year-old individuals is -0.7 pp (p $=$ 0.05) over one year, -0.9 pp (p $=$ 0.04) over three years, and -0.7 pp (p $=$ 0.16) over five years, and among individuals aged 65 or more -0.9 pp (p $=$ 0.04) over one year, -1.0 pp (p $=$ 0.04) over three years, and -1.0 pp (p $=$ 0.06) over five years after its introduction. The corresponding estimates for men and the remaining age groups were not significant at the 5\% level. Overall, the results suggest that the reduction in the smoking rate is driven by women and individuals aged 25--44 and 65+. 

The conditional common trends assumption is also assessed across all subgroups. In the same fashion as for previous results, \cref{tab:bb:hetRes} reports placebo test results. In all subgroups except men, the pretreatment effect in the year directly before implementation is not significant at the 5\% level. Over the five- and ten-year pretreatment windows, the average year-specific ATETs are insignificant across all subgroups at the 5\% level. The conditional common trends assumption generally holds, except for men in the year immediately preceding the billboard ban implementation. Nonetheless, this identifying assumption never holds across all treatment groups introducing the billboard ban at different points in time, just like in the analysis based on the whole sample.

\Cref{fig:bb:ifeRes} shows the same event-study estimates, stratified by gender and age, based on the IFEct estimator by \textcite{Liu24}. The IFEct estimates confirm that the reduction in smoking rates would be driven by women and individuals aged 25--44 and 65+. For women, the year-specific ATETs range from -0.5 to -0.6 pp. The year-specific ATETs range from -0.7 to -1.2 pp for individuals aged 25--44 and from -0.5 to -0.9 pp for individuals aged 65+. In each subgroup, all year-specific ATETs in posttreatment periods are negative and most often significant at the 5\% level. The first two year-specific ATETs after billboard ban implementation are negative and significant at the 5\% level in each subgroup.

The pretreatment year-specific ATETs are close to zero and most often not significant at the 5\% level across all subgroups, except for the 15--24-year-old individuals. This indicates that the latent factor model manages to adjust for unobserved time-varying confounding in most cases. However, for the individuals aged 15--24, the IFEct approach did not perform well, having pretreatment ATETs ranging from relatively low to high estimates. Generally, the pretreatment ATETs support the validity of the IFEct estimates, except for the age group 15--24 where validity is limited.

Concluding, both estimation approaches suggest that women and individuals aged 25--44 and 65+ are the drivers of the reduction in smoking rates following the billboard ban. The staggered DiD approach does not find significant reductions at the 5\% level in the remaining subgroups. In contrast, the IFEct estimator finds significant reductions across all subgroups. Again, both estimators should be understood as complementary, given their differing identification assumptions. Women, individuals aged 25--44, and those aged 65 and older are consistently the drivers of the reduction in smoking rates across both estimation approaches. Overall, our findings thus suggest that the reduction in smoking rates is driven by these three subgroups.

\section{Robustness Checks}
\label{sec:bb:robustChecks}

We perform a series of robustness checks to evaluate the sensitivity of the main estimates. These estimates are based on the staggered DiD estimator by \textcite{Callaway21}, applying the main specification while varying specific parameters. The robustness checks include varying the covariate composition, the units used as the control group, the retrospective reconstruction of the smoking status, long-gap estimation of ATETs in the pretreatment periods, and specifications accounting for spillover effects. \Cref{tab:bb:mainRes} presents the estimates of the robustness checks, including the main estimate in column (4).

First, we explore how sensitive the estimates are with respect to changes in the covariate composition. For this purpose, the analysis is performed once without covariates in column (1), with only gender and age covariates in column (2), and with only sales and smoking ban covariates in column (3). The main estimates in column (4) are based on all covariates included without interactions. The average year-specific ATETs using the main specification were -0.3 pp (p $=$ 0.11) over one year, -0.5 pp (p $=$ 0.02) over three years, and -0.4 pp (p $=$ 0.08) over five years after billboard ban introduction. Without covariates, these estimates are close to zero and not significant at the 5\% level. Including only the socio-demographic covariates, i.e., gender and age, yields similar estimates close to zero which are again not significant at the 5\% level. Including only the tobacco prevention policy covariates, i.e., the sales and smoking bans, in contrast, produces estimates ranging between -0.4 and -0.7 and which are all significant at the 5\% level. The estimates appear to be particularly sensitive to the inclusion of tobacco prevention policies. Once these policies are added as controls, the estimated effects shift from near zero to more negative values.

We further estimate a model specification that interacts all covariates, i.e., age, gender, sales ban, and smoking ban, as reported in column (5). This model is more flexible than the main specification as it allows for interactions. The resulting average year-specific ATETs following the introduction of the billboard ban are -0.3 pp (p $=$ 0.15) over one year, -0.7 pp (p $<$ 0.01) over three years, and -0.8 pp (p $<$ 0.01) over five years after its adoption. Relative to the main estimates, the effects become more pronounced, particularly over three and five years after the billboard ban. Importantly, standard errors remain comparable to those in the main specification, suggesting that the added flexibility does not substantially increase estimation noise. These results are consistent with the main findings and indicate potentially stronger reductions in the smoking rate when allowing for covariate interactions. 

Second, we use never-treated units as the control group instead of the not-yet-treated units. The not-yet-treated units were used in the main specification because they are likely more comparable to treated units and because the number of never-treated units is relatively small, which could reduce statistical power. However, not-yet-treated units may be more susceptible to bias from anticipation effects, particularly if the billboard ban was publicly discussed prior to implementation. In contrast, never-treated units are less prone to such bias. The average year-specific ATETs using never-treated units as a control group are presented with all covariates included in column (6), and with all covariates interacted in column (7). Compared to the main estimates, both specifications yield slightly stronger reductions in the smoking rate and higher significance levels. Standard errors remain similar in comparison to the main specification. These findings indicate that the estimates are not sensitive to the choice of control group.

Third, we evaluate the sensitivity of the estimates with respect to the retrospective reconstruction of smoking status. Each individual is observed only once at the survey date, and a panel is constructed by retrospectively calculating their smoking history, as described in \cref{sec:bb:swissHealthSurvey}. For that reason, we only observe the canton of residence at the survey date, while prior residency is unknown. Consequently, spillover effects may arise if individuals moved between cantons with and without billboard bans. To assess the sensitivity to such mobility, we restrict the reconstructed smoking histories to 5, 10, and 15 years prior to the survey date. The corresponding estimates are reported in columns (8)--(10). Across all three specifications, the average year-specific ATETs are comparable to the main estimates, i.e., negative throughout and several times significant at the 5\% level. These findings suggest that the estimates are not sensitive to the length of the reconstructed smoking histories and that mobility is not a relevant constraint for this analysis. Moreover, this exercise provides a check on potential recall bias among participants. Such bias may arise if individuals remember their more recent smoking history more accurately than more distant behavior. The consistency of the results when restricting the analysis to smoking histories from 5, 10, and 15 years prior to the survey suggests that recall bias is likely to be limited.

Fourth, the event-study estimates of the main specification in \cref{fig:bb:mainResNotYet} are asynchronous. While the year-specific ATETs use the period immediately preceding the introduction of the billboard ban as the reference period, the year-specific ATETs in the pretreatment periods compare one year to two years before adoption, two years to three years before adoption, and so on. Hence, the reference period changes across pretreatment periods when using short-gap estimation. For completeness, \cref{fig:bb:mainResNotYet_long} shows the main specification using long-gap estimation instead of short-gap estimation for the ATETs in the pretreatment periods. Under this approach, all ATETs are estimated using a single reference period, namely the period immediately preceding the implementation of the billboard ban. By construction, the ATETs in the posttreatment periods are unaffected by this change. The ATETs in the pretreatment periods do change, but for the most part their confidence intervals include zero. However, the placebo test across all treatment groups is rejected under long-gap estimation, as is the case for short-gap estimation.

Fifth, mobility across cantons may bias the results. In particular, individuals may reside in one canton and work in another. As a result, individuals living in a canton without a billboard ban may still be exposed to such a ban through commuting, and vice versa. As discussed earlier in \cref{sec:bb:empStrategy}, this is likely to attenuate the estimated reduction in smoking due to billboard bans. \Cref{fig:bb:mainResSpill} presents event-study estimates that control for such spillover effects. Starting from the main specification, a dummy variable indicating whether a neighboring canton has already introduced a billboard ban is included. The average ATET across all posttreatment periods remains a reduction of 0.4 pp (p $>$ 0.05). However, when this dummy is interacted with the share of commuters who travel outside their canton of residence, the estimated average reduction increases to 2.0 pp (p $<$ 0.01). The commuter share is constructed as the ratio of the number of out-of-canton commuters to the cantonal population, based on 2010 data from the \textcite{BFS24}, the earliest available year. Shares range from 1\% to 33\%, indicating substantial regional differences. This latter specification better captures actual exposure to spillovers, as it accounts for the effective share of commuting individuals. As expected, the main specification provides a conservative estimate and likely understates the true effect of billboard bans.

In summary, estimates are sensitive to the inclusion of other tobacco prevention policies, while socio-demographic factors have little influence. This underscores the importance of controlling for confounding from these policies. The estimates are robust to the choice of control group, whether using not-yet-treated or never-treated units. Allowing for interactions between covariates yields more pronounced reductions in smoking rates, without considerably increasing standard errors. Regarding data limitations, restricting the retrospective reconstruction of smoking histories suggests that unobserved residential mobility does not materially bias the results. Moreover, results adjusting for spillover effects indicate that these attenuate the estimated treatment effect. Overall, the main findings using staggered DiD are robust to a range of model specifications and sensitivity checks.

Compared to the staggered DiD estimator, conducting robustness checks with the IFEct estimator is more challenging, as it requires a sufficient number of observations to learn the latent factor structure. As a result, analyses based on shorter smoking-history windows are limited. Moreover, because counterfactual outcomes are imputed, there is no variation from not-yet-treated or never-treated units as in staggered DiD. Robustness to the choice of covariates can, however, be assessed by estimating the model with no covariates, only socio-demographic covariates, and only policy covariates. \Cref{fig:bb:ifeResCovs} presents the corresponding event-study estimates. Interestingly, the results are very similar across all three specifications. Regardless of the covariates included, the estimates indicate an immediate reduction in the smoking rate of about 1.0 pp that persists over time and is statistically significant at the 5\% level. This suggests that the latent factor model already captures the most important confounding factors and is therefore relatively insensitive to the inclusion or exclusion of these covariates.

\section{Conclusion}
\label{sec:bb:conclusion}

This study provides evidence on the effectiveness of tobacco advertising bans, specifically tobacco billboard bans. The main results suggest that tobacco billboard bans reduce smoking rates. The IFEct estimator finds a sustained decline in the smoking rate of 0.9 pp (p $<$ 0.01), while staggered DiD suggests a more moderate average reduction of 0.4 pp (p $=$ 0.08) over five years after the tobacco billboard ban implementation. Reductions of up to 0.9 pp correspond to a relative decline of about 3\%, given an average smoking rate of 31\% in the sample. This effect is equivalent to approximately a 10\% to 30\% increase in tobacco prices, based on estimates of the participation price elasticity of smoking ranging from 0.1 to 0.3 \parencite{DeCicca22}.

The literature on tobacco advertising remains inconclusive regarding its role in smoking. Some studies suggest that tobacco advertising initiates smoking among non-smokers or increases consumption among current smokers, while others do not find such market expansion effects. We provide new evidence on this matter by exploiting a quasi-experimental setting and applying state-of-the-art methods for causal inference, both important improvements over existing work. Our findings support the market expansion hypothesis, providing a rationale for advertising regulation to promote public health. Overall, our results suggest that advertising bans can be relevant policy tools, complementing other tobacco prevention policies such as taxation and place-based smoking bans.

Heterogeneity analysis across gender and age shows that the reduction in smoking is driven by women and individuals aged 25--44 and 65+. The absence of effectiveness among youth aged 15--24 is surprising, as advertising bans are commonly motivated by the goal of preventing youth smoking initiation. Three tentative reasons may help explain this puzzle. First, other factors may be more relevant for this age group, such as tobacco prices and social influences. Second, while substitution from billboards to other advertising channels appears limited, such substitution might nevertheless have occurred following the billboard bans. As a result, the most affected cantons could have been targeted through alternative channels, likely focusing on the youth population and offsetting the effect of the ban among them. Third, the effect is driven by women and persons aged 25--44, a subpopulation for whom quitting smoking may be especially important, given the higher prevalence of pregnancy during these years. The absence of advertising may mitigate self-control problems, reduce temptations, and thereby support smoking cessation attempts. These interpretations remain speculative, and further research is warranted to better understand the underlying mechanisms.

\section*{Acknowledgments}

Thanks are expressed to Martin Huber, Marc Suhrcke, Michela Bia, Igor Franceti{\'c}, Jannis St{\"o}ckel, Christian Saenz, as well as participants at the Symposium of Causal Inference in the Health Sciences, Workshop of the Swiss Network on Public Economics, Swiss Public Health Conference, Society for Longitudinal and Lifecourse Studies Conference, seminar of the Lausanne Center for Health Economics, Behavior, and Policy, seminar of the Luxembourg Institute of Socio-Economic Research, seminar of the University of St.\ Gallen, dgg{\"o} Annual Conference, Swiss Health Economics Workshop, IRDES-LIRAES Workshop on Applied Health Economics and Policy Evaluation, Tobacco Online Policy Seminar, and internal seminar of our Department of Economics, for their valuable comments. Jeremy Proz is thanked for his research assistance. Financial support from the Swiss Tobacco Prevention Fund is gratefully acknowledged.

\section*{Declaration of Generative AI and AI-assisted Technologies in the Writing Process}

During the preparation of this work the authors used Microsoft Copilot in order to rephrase or partially generate text, as well as to refine and partially generate code. After using this tool, the authors reviewed and edited the content as needed and take full responsibility for the content of the publication.

\section*{References}

\printbibliography[heading=none]

@Article{Abadie05,
  author  = {Abadie, Alberto},
  title   = {Semiparametric Difference-in-Differences Estimators},
  doi     = {10.1111/0034-6527.00321},
  number  = {1},
  pages   = {1--19},
  volume  = {72},
  journal = {The Review of Economic Studies},
  year    = {2005},
}

@Article{Borusyak24,
  author  = {Borusyak, Kirill and Jaravel, Xavier and Spiess, Jann},
  title   = {Revisiting Event-Study Designs: Robust and Efficient Estimation},
  doi     = {10.1093/restud/rdae007},
  number  = {6},
  pages   = {3253--3285},
  volume  = {91},
  journal = {The Review of Economic Studies},
  year    = {2024},
}

@Article{Callaway21,
  author  = {Callaway, Brantly and Sant’Anna, Pedro H. C.},
  title   = {Difference-in-Differences With Multiple Time Periods},
  doi     = {10.1016/j.jeconom.2020.12.001},
  number  = {2},
  pages   = {200--230},
  volume  = {225},
  journal = {Journal of Econometrics},
  year    = {2021},
}

@Article{DeChaisemartin20,
  author  = {{De Chaisemartin}, Cl{\'e}ment and d’Haultfoeuille, Xavier},
  title   = {Two-Way Fixed Effects Estimators With Heterogeneous Treatment Effects},
  doi     = {10.1257/aer.20181169},
  number  = {9},
  pages   = {2964--2996},
  volume  = {110},
  journal = {American Economic Review},
  year    = {2020},
}

@Article{Gallet03,
  author  = {Gallet, Craig A. and List, John A.},
  title   = {Cigarette Demand: A Meta-Analysis of Elasticities},
  doi     = {10.1002/hec.765},
  number  = {10},
  pages   = {821--835},
  volume  = {12},
  journal = {Health Economics},
  year    = {2003},
}

@Article{Gobillon16,
  author  = {Gobillon, Laurent and Magnac, Thierry},
  title   = {Regional Policy Evaluation: Interactive Fixed Effects and Synthetic Controls},
  doi     = {10.1162/REST_a_00537},
  number  = {3},
  pages   = {535--551},
  volume  = {98},
  journal = {Review of Economics and Statistics},
  year    = {2016},
}

@Article{Goel06,
  author  = {Goel, Rajeev K. and Nelson, Michael A.},
  title   = {The Effectiveness of Anti-Smoking Legislation: A Review},
  doi     = {10.1111/j.0950-0804.2006.00282.x},
  number  = {3},
  pages   = {325--355},
  volume  = {20},
  journal = {Journal of Economic Surveys},
  year    = {2006},
}

@Article{GoodmanBacon21,
  author  = {Goodman-Bacon, Andrew},
  title   = {Difference-in-Differences With Variation in Treatment Timing},
  doi     = {10.1016/j.jeconom.2021.03.014},
  number  = {2},
  pages   = {254--277},
  volume  = {225},
  journal = {Journal of Econometrics},
  year    = {2021},
}

@Article{Heckman08,
  author  = {Heckman, James J. and Flyer, Fredrick and Loughlin, Colleen},
  title   = {An Assessment of Causal Inference in Smoking Initiation Research and a Framework for Future Research},
  doi     = {10.1111/j.1465-7295.2007.00078.x},
  number  = {1},
  pages   = {37--44},
  volume  = {46},
  journal = {Economic Inquiry},
  year    = {2008},
}

@Article{Lechner11,
  author  = {Lechner, Michael},
  title   = {The Estimation of Causal Effects by Difference-in-Difference Methods},
  doi     = {10.1561/0800000014},
  number  = {3},
  pages   = {165--224},
  volume  = {4},
  journal = {Foundations and Trends® in Econometrics},
  year    = {2011},
}

@Article{Liu24,
  author  = {Liu, Licheng and Wang, Ye and Xu, Yiqing},
  title   = {A Practical Guide to Counterfactual Estimators for Causal Inference With Time-Series Cross-Sectional Data},
  doi     = {10.1111/ajps.12723},
  number  = {1},
  pages   = {160--176},
  volume  = {68},
  journal = {American Journal of Political Science},
  year    = {2024},
}

@Article{Murray20,
  author  = {Murray, Christopher J. L. and Aravkin, Aleksandr Y. and Zheng, Peng and Abbafati, Cristiana and Abbas, Kaja M. and Abbasi-Kangevari, Mohsen and Abd-Allah, Foad and Abdelalim, Ahmed and Abdollahi, Mohammad and Abdollahpour, Ibrahim and others},
  title   = {Global Burden of 87 Risk Factors in 204 Countries and Territories, 1990--2019: A Systematic Analysis for the Global Burden of Disease Study 2019},
  doi     = {10.1016/S0140-6736(20)30752-2},
  number  = {10258},
  pages   = {1223--1249},
  volume  = {396},
  journal = {The Lancet},
  year    = {2020},
}

@Article{Nelson10,
  author  = {Nelson, Jon P.},
  title   = {What Is Learned From Longitudinal Studies of Advertising and Youth Drinking and Smoking? A Critical Assessment},
  doi     = {10.3390/ijerph7030870},
  number  = {3},
  pages   = {870--926},
  volume  = {7},
  journal = {International Journal of Environmental Research and Public Health},
  year    = {2010},
}

@Article{Neyman23,
  author  = {Neyman, Jerzy},
  title   = {On the Application of Probability Theory to Agricultural Experiments. Essay on Principles},
  pages   = {1--51},
  journal = {Annals of Agricultural Sciences},
  year    = {1923},
}

@Article{Qi13,
  author  = {Qi, Shi},
  title   = {The Impact of Advertising Regulation on Industry: The Cigarette Advertising Ban of 1971},
  doi     = {10.1111/1756-2171.12018},
  number  = {2},
  pages   = {215--248},
  volume  = {44},
  journal = {The RAND Journal of Economics},
  year    = {2013},
}

@Article{Reitsma21,
  author  = {Reitsma, Marissa B. and Kendrick, Parkes J. and Ababneh, Emad and Abbafati, Cristiana and Abbasi-Kangevari, Mohsen and Abdoli, Amir and Abedi, Aidin and Abhilash, E.S. and Abila, Derrick Bary and Aboyans, Victor and others},
  title   = {Spatial, Temporal, and Demographic Patterns in Prevalence of Smoking Tobacco Use and Attributable Disease Burden in 204 Countries and Territories, 1990--2019: A Systematic Analysis From the Global Burden of Disease Study 2019},
  doi     = {10.1016/S0140-6736(21)01169-7},
  number  = {10292},
  pages   = {2337--2360},
  volume  = {397},
  journal = {The Lancet},
  year    = {2021},
}

@Article{Rubin74,
  author  = {Rubin, Donald B.},
  title   = {Estimating Causal Effects of Treatments in Randomized and Nonrandomized Studies},
  doi     = {10.1037/h0037350},
  number  = {5},
  pages   = {688--701},
  volume  = {66},
  journal = {Journal of Educational Psychology},
  year    = {1974},
}

@Article{Saffer00,
  author  = {Saffer, Henry and Chaloupka, Frank},
  title   = {The Effect of Tobacco Advertising Bans on Tobacco Consumption},
  doi     = {10.1016/S0167-6296(00)00054-0},
  number  = {6},
  pages   = {1117--1137},
  volume  = {19},
  journal = {Journal of Health Economics},
  year    = {2000},
}

@Article{Sun21,
  author  = {Sun, Liyang and Abraham, Sarah},
  title   = {Estimating Dynamic Treatment Effects in Event Studies With Heterogeneous Treatment Effects},
  doi     = {10.1016/j.jeconom.2020.09.006},
  number  = {2},
  pages   = {175--199},
  volume  = {225},
  journal = {Journal of Econometrics},
  year    = {2021},
}

@Article{Xu17,
  author  = {Xu, Yiqing},
  title   = {Generalized Synthetic Control Method: Causal Inference With Interactive Fixed Effects Models},
  doi     = {10.1017/pan.2016.2},
  number  = {1},
  pages   = {57--76},
  volume  = {25},
  journal = {Political Analysis},
  year    = {2017},
}

@Report{BAG15,
  author      = {{Federal Office of Public Health}},
  institution = {Federal Office of Public Health},
  title       = {Basisinformation zur Tabakwerbung},
  url         = {https://www.bag.admin.ch/dam/bag/de/dokumente/npp/tabak/basisinformation-tabakwerbung.pdf.download.pdf/2015_Basisinformation_Tabakwerbung_DE.pdf},
  address     = {Bern},
  year        = {2015},
}

@misc{BAG25swiss,
  author       = {{Federal Office of Public Health}},
  title        = {Gesetzgebung Tabak und elektronische Zigaretten},
  year         = {2025},
  url          = {https://www.bag.admin.ch/de/gesetzgebung-tabak-und-elektronische-zigaretten},
  note         = {Accessed: 2025-10-18}
}

@misc{BAG25cantonal,
  author       = {{Federal Office of Public Health}},
  title        = {Tabakpolitik in den Kantonen},
  year         = {2025},
  url          = {https://www.bag.admin.ch/de/tabakpolitik-in-den-kantonen},
  note         = {Accessed: 2025-10-18}
}

@misc{Rios23,
  title={CSDID: Stata Module for the Estimation of Difference-in-Difference Models With Multiple Time Periods},
  author={Rios-Avila, Fernando and Sant'Anna, Pedro H. C. and Callaway, Brantly},
  year={2023},
}

@Misc{SGB9717,
  author    = {{Federal Statistical Office}},
  title     = {Swiss Health Survey, 1997--2017},
  address   = {Neuch{\^a}tel},
  publisher = {Federal Statistical Office},
  year      = {2024},
}

@Article{DeCicca22,
  author  = {DeCicca, Philip and Kenkel, Donald and Lovenheim, Michael F.},
  date    = {2022},
  title   = {The Economics of Tobacco Regulation: A Comprehensive Review},
  doi     = {10.1257/jel.20201482},
  number  = {3},
  pages   = {883–-970},
  volume  = {60},
  journal = {Journal of Economic Literature},
}

@Misc{DeChaisemartin23,
  author = {{De Chaisemartin}, Cl{\'e}ment and D'Haultf{\oe}uille, Xavier},
  title  = {Credible Answers to Hard Questions: Differences-in-Differences for Natural Experiments},
  doi    = {10.2139/ssrn.4487202},
  note   = {Available at SSRN 4487202},
  year   = {2023},
}

@Article{Marti14,
  author  = {Marti, Joachim},
  title   = {The Impact of Tobacco Control Expenditures on Smoking Initiation and Cessation},
  doi     = {10.1002/hec.2993},
  number  = {12},
  pages   = {1397--1410},
  volume  = {23},
  journal = {Health Economics},
  year    = {2014},
}

@Misc{BFS24,
  author    = {{Federal Statistical Office}},
  title     = {St{\"a}ndige Wohnbev{\"o}lkerung, Erwerbst{\"a}tige, Auszubildende: Pendler/-innen nach Kantonen (T 11.4.4.1)},
  note      = {PEND (Pendlermobilit{\"a}t) \& Strukturerhebung (SE)},
  url       = {https://www.bfs.admin.ch/bfsstatic/dam/assets/1862650/master},
  address   = {Neuch{\^a}tel},
  publisher = {Federal Statistical Office},
  year      = {2024},
}

@InCollection{Kuenzler18,
  author    = {Kuenzler, Johanna},
  booktitle = {Learning in Public Policy: Analysis, Modes and Outcomes},
  title     = {The Hard Case for Learning: Explaining the Diversity of Swiss Tobacco Advertisement Bans},
  doi       = {10.1007/978-3-319-76210-4_12},
  pages     = {267--294},
  publisher = {Springer},
  year      = {2018},
}

@Misc{BAZG26,
  author = {{Federal Office for Customs and Border Security}},
  title  = {Zigarettenpreis je P{\"a}ckchen und Tabaksteueranteil in CHF, 2005--2025},
  note   = {Accessed: 2026-07-01},
  url    = {https://www.bazg.admin.ch/dam/de/sd-web/Zano59keL6Iw/Tabelle%20von%20Grafik%20Preis-%20und%20Tabaksteuerentwicklung%20in%20der%20CH%20d%20bis%202025.pdf},
  year   = {2026},
}

@Misc{ATSchweiz26,
  author = {{Swiss Association for Tobacco Control}},
  title  = {Prices for Tobacco and Nicotine Products in Switzerland},
  note   = {Accessed: 2026-07-01},
  url    = {https://www.at-schweiz.ch/en/advocacy/prices-taxes/preise-fur-tabak-und-nikotinprodukte-in-der-schweiz/},
  year   = {2022},
}


\newpage

\begin{appendices}
	
\section{Annex}

\begin{table}[htbp!]
	\caption{Comparison of sample composition before and after panel reconstruction, by gender and age} 
	\label{tab:bb:comparisonComposition}
	\centering
	\scriptsize
	\begin{tabular}{p{.05\textwidth}p{.15\textwidth}p{.075\textwidth}p{.075\textwidth}p{.075\textwidth}p{.075\textwidth}p{.075\textwidth}p{.075\textwidth}}  
		 & & \multicolumn{2}{c}{Gender} & \multicolumn{4}{c}{Age} \\       
		\cmidrule(lr){3-4} \cmidrule(lr){5-8}
		Year & Data & Women & Men & 15--24 & 25--44 & 45--64 & 65+ \\
		\hline
		1997 & Original & 0.56 & 0.44 & 0.11 & 0.42 & 0.28 & 0.19 \\ 
		 & Panel & 0.55 & 0.45 & 0.15 & 0.43 & 0.30 & 0.11 \\ 
		 & Difference (in \%) & -1.28 & 1.65 & 38.57 & 4.27 & 7.54 & -42.38 \\ 
		2002 & Original & 0.56 & 0.44 & 0.09 & 0.37 & 0.32 & 0.22 \\ 
		 & Panel & 0.55 & 0.45 & 0.13 & 0.41 & 0.32 & 0.14 \\ 
		 & Difference (in \%) & -1.04 & 1.31 & 50.35 & 10.42 & -1.87 & -34.61 \\ 
		2007 & Original & 0.56 & 0.44 & 0.10 & 0.34 & 0.33 & 0.24 \\ 
		 & Panel & 0.55 & 0.45 & 0.13 & 0.37 & 0.33 & 0.17 \\ 
		 & Difference (in \%) & -1.86 & 2.35 & 35.74 & 10.17 & 0.40 & -29.57 \\ 
		2012 & Original & 0.53 & 0.47 & 0.14 & 0.29 & 0.35 & 0.22 \\ 
		 & Panel & 0.53 & 0.47 & 0.14 & 0.31 & 0.36 & 0.20 \\ 
		 & Difference (in \%) & 0.88 & -0.98 & -1.87 & 6.63 & 1.19 & -9.34 \\ 
		2017 & Original & 0.53 & 0.47 & 0.13 & 0.28 & 0.36 & 0.23 \\ 
		 & Panel & 0.53 & 0.47 & 0.13 & 0.28 & 0.36 & 0.23 \\ 
		 & Difference (in \%) & 0.00 & 0.00 & 0.00 & 0.00 & 0.00 & 0.00 \\
		\hline\hline
	\end{tabular}
	\caption*{\footnotesize \textit{Note: This table reports the share of subgroups by gender (women and men) and age (15--24, 25--44, 45--64, and 65+ years old) in the original cross-sectional data and the reconstructed panel data. The third row for each year shows the relative difference (in percent) between the data before and after panel reconstruction.}}
\end{table}

\FloatBarrier

\newpage

\begin{table}[htbp!]
	\caption{Comparison of smoking rates before and after panel reconstruction, by gender and age} 
	\label{tab:bb:comparisonSmoking}
	\centering
	\scriptsize
	\begin{tabular}{p{.05\textwidth}p{.15\textwidth}p{.075\textwidth}p{.075\textwidth}p{.075\textwidth}p{.075\textwidth}p{.075\textwidth}p{.075\textwidth}p{.075\textwidth}}  
		& & & \multicolumn{2}{c}{Gender} & \multicolumn{4}{c}{Age} \\ 
		\cmidrule(lr){4-5} \cmidrule(lr){6-9}    
		Year & Data & Total & Women & Men & 15--24 & 25--44 & 45--64 & 65+ \\
		\hline
		1997 & Original & 0.34 & 0.29 & 0.40 & 0.45 & 0.42 & 0.32 & 0.15 \\ 
		 & Panel & 0.33 & 0.29 & 0.38 & 0.34 & 0.40 & 0.29 & 0.13 \\ 
		 & Difference (in \%) & -3.62 & -2.74 & -4.85 & -26.09 & -3.17 & -7.82 & -12.73 \\ 
		2002 & Original & 0.30 & 0.27 & 0.35 & 0.38 & 0.38 & 0.31 & 0.14 \\ 
		 & Panel & 0.32 & 0.28 & 0.37 & 0.36 & 0.38 & 0.31 & 0.13 \\ 
		 & Difference (in \%) & 4.52 & 3.38 & 5.28 & -3.10 & 0.52 & -2.13 & -8.54 \\ 
		2007 & Original & 0.27 & 0.23 & 0.31 & 0.34 & 0.33 & 0.29 & 0.13 \\ 
		 & Panel & 0.30 & 0.26 & 0.34 & 0.34 & 0.35 & 0.31 & 0.13 \\ 
		 & Difference (in \%) & 9.98 & 10.04 & 9.28 & -2.34 & 7.15 & 5.64 & -2.27 \\ 
		2012 & Original & 0.28 & 0.25 & 0.31 & 0.36 & 0.33 & 0.29 & 0.14 \\ 
		 & Panel & 0.27 & 0.24 & 0.31 & 0.33 & 0.33 & 0.28 & 0.13 \\ 
		 & Difference (in \%) & -1.50 & -2.34 & -0.55 & -6.11 & -0.42 & -2.86 & -5.74 \\ 
		2017 & Original & 0.26 & 0.23 & 0.30 & 0.32 & 0.33 & 0.26 & 0.14 \\ 
		 & Panel & 0.26 & 0.23 & 0.30 & 0.32 & 0.33 & 0.26 & 0.14 \\ 
		 & Difference (in \%) & 0.00 & 0.00 & 0.00 & 0.00 & 0.00 & 0.00 & 0.00 \\
		\hline\hline
	\end{tabular}
	\caption*{\footnotesize \textit{Note: This table reports the smoking rate of subgroups by gender (women and men) and age (15--24, 25--44, 45--64, and 65+ years old) in the original cross-sectional data and the reconstructed panel data. The third row for each year shows the relative difference (in percent) between the data before and after panel reconstruction.}}
\end{table}

\FloatBarrier

\newpage

\begin{table}[htbp!]
	\caption{Introduction of cantonal tobacco billboard, sales, and smoking bans (as of December 31, 2017)} 
	\label{tab:bb:cantonalPolicies}
	\centering
	\scriptsize
	\begin{tabular}{p{.25\textwidth}p{.15\textwidth}p{.15\textwidth}p{.15\textwidth}}           
		Canton & Billboard ban & Sales ban & Smoking ban \\
		\hline
		Aargau (AG) &  & 2010-01-01 & 2010-05-01 \\ 
		Appenzell Ausserrhoden (AR) & 2008-01-01 & 2008-01-01 & 2010-05-01 \\ 
		Appenzell Innerrhoden (AI) &  &  & 2010-05-01 \\ 
		Basel-Landschaft (BL) & 2007-01-01 & 2007-01-01 & 2010-05-01 \\ 
		Basel-Stadt (BS) & 1997-02-06 & 2007-08-01 & 2010-04-01 \\ 
		Bern (BE) & 2007-01-01 & 2007-01-01 & 2009-07-01 \\ 
		Fribourg (FR) &  & 2009-01-01 & 2010-01-01 \\ 
		Geneva (GE) & 2000-10-20 & & 2009-11-01 \\ 
		Glarus (GL) &  & 2014-01-01 & 2010-05-01 \\ 
		Grisons (GR) & 2006-07-01 & 2006-07-01 & 2008-03-01 \\ 
		Jura (JU) &  & 2013-01-01 & 2010-05-01 \\ 
		Lucerne (LU) &  & 2006-01-01 & 2010-05-01 \\ 
		Neuch\^atel (NE) &  & 2015-01-01 & 2009-04-01 \\ 
		Nidwalden (NW) &  & 2009-03-01 & 2010-05-01 \\ 
		Obwalden (OW) & 2016-02-01 & 2016-02-01 & 2010-05-01 \\ 
		Schaffhausen (SH) &  & 2013-01-01 & 2010-05-01 \\ 
		Schwyz (SZ) &  &  & 2010-05-01 \\ 
		Solothurn (SO) & 2007-07-01 & 2007-01-01 & 2009-01-01 \\ 
		St. Gallen (SG) & 2006-10-01 & 2006-10-01 & 2010-05-01 \\ 
		Thurgau (TG) & 2007-01-01 & 2007-01-01 & 2010-05-01 \\ 
		Ticino (TI) & 2009-05-01 & 2013-09-01 & 2007-04-12 \\ 
		Uri (UR) & 2009-09-01 & 2009-09-01 & 2009-09-01 \\ 
		Valais (VS) & 2009-07-01 & 2008-01-01 & 2009-07-01 \\ 
		Vaud (VD) & 2007-07-01 & 2006-01-01 & 2009-09-15 \\ 
		Zug (ZG) & 2010-03-01 & 2010-03-01 & 2010-03-01 \\ 
		Zurich (ZH) & 2008-07-01 & 2008-07-01 & 2010-05-01 \\
		\hline\hline
	\end{tabular}
	\caption*{\footnotesize \textit{Note: This table reports the implementation dates of the cantonal policies, formatted as YYYY-MM-DD. The billboard ban refers to the first introduction of a tobacco billboard ban in public areas or in areas visible from public areas. The sales ban indicates the earliest introduction of a tobacco sales ban for minors under the age of 16 or 18. The smoking ban denotes the first introduction of smoking bans in public areas.}}
\end{table}

\FloatBarrier

\newpage

\begin{table}[htbp!]
	\caption{Average year-specific effects of tobacco billboard bans on the smoking rate} 
	\label{tab:bb:mainRes}
	\centering
	\scriptsize
	\begin{tabular}{p{.10\textwidth}p{.06\textwidth}p{.06\textwidth}p{.06\textwidth}p{.06\textwidth}p{.06\textwidth}p{.06\textwidth}p{.06\textwidth}p{.06\textwidth}p{.06\textwidth}p{.06\textwidth}}           
	& \multicolumn{5}{c}{Varying covariate compositions} & \multicolumn{2}{c}{Never-treated} & \multicolumn{3}{c}{5-, 10-, and 15-year histories} \\          
	\cmidrule(lr){2-6} \cmidrule(lr){7-8} \cmidrule(lr){9-11}
	&       (1)&      (2)&      (3)&       (4)&   (5)&     (6) &    (7)&    (8) & (9) & (10)\\
	\hline
	\multicolumn{11}{l}{Average year-specific posttreatment effect} \\
	\hline
	\multicolumn{11}{l}{0--1 years after billboard ban introduction} \\ 
	ATET &      -0.0004&       0.0002&       -0.0040&      -0.0029&      -0.0028&      -0.0037&      -0.0038&      -0.0007&      -0.0051&      -0.0033\\
	Std. error          &       0.0014&       0.0014&       0.0018&       0.0018&       0.0019&       0.0019&       0.0017&       0.0049&       0.0027&       0.0019\\
	P-value     &       0.7956&       0.8966&       0.0232&       0.1098&       0.1494&        0.0440&        0.0290&       0.8811&       0.0586&       0.0789\\
	\multicolumn{11}{l}{0--3 years after billboard ban introduction} \\ 
	ATET &      -0.0005&       0.0003&      -0.0071&      -0.0052&      -0.0068&      -0.0062&      -0.0072&      -0.0083&      -0.0078&      -0.0057\\
	Std. error          &       0.0019&       0.0019&       0.0021&       0.0022&       0.0026&       0.0022&       0.0024&       0.0046&       0.0032&       0.0023\\
	P-value     &       0.7926&       0.8552&        0.0010&       0.0177&       0.0083&       0.0049&       0.0028&       0.0709&        0.0150&       0.0137\\
	\multicolumn{11}{l}{0--5 years after billboard ban introduction} \\ 
	ATET &      -0.0003&       0.0008&      -0.0066&      -0.0043&      -0.0077&      -0.0054&      -0.0081&      -0.0071&      -0.0068&      -0.0047\\
	Std. error           &       0.0022&       0.0022&       0.0024&       0.0024&       0.0029&       0.0024&       0.0029&        0.0050&       0.0034&       0.0025\\
	P-value     &       0.9021&       0.7131&       0.0062&       0.0772&       0.0088&       0.0252&       0.0048&       0.1527&       0.0497&       0.0649\\
	\hline
	\multicolumn{11}{l}{Average year-specific pretreatment effect} \\
	\hline
	\multicolumn{11}{l}{1 year before billboard ban introduction} \\ 
	Coefficient &       0.0017&       0.0013&       0.0003&       0.0007&       0.0009&      -0.0013&      -0.0011&       0.0011&      -0.0004&       0.0008\\
	Std. error          &       0.0011&       0.0011&       0.0012&       0.0012&       0.0012&       0.0012&       0.0012&       0.0029&       0.0017&       0.0013\\
	P-value     &       0.1004&       0.2108&       0.7873&       0.5618&       0.4547&       0.2748&       0.3365&       0.7081&       0.8006&       0.5159\\ 
	\multicolumn{11}{l}{1--5 years before billboard ban introduction} \\ 
	Coefficient &       0.0007&       0.0005&           0.0000&      -0.0001&           0.0000&           0.0000&       0.0001&      -0.0001&      -0.0001&      -0.0003\\
	Std. error          &       0.0004&       0.0004&       0.0004&       0.0004&       0.0004&       0.0005&       0.0005&       0.0015&       0.0007&       0.0005\\
	P-value     &       0.1039&       0.1933&       0.9388&       0.8354&        0.9040&       0.9633&       0.9153&       0.9394&       0.8441&       0.5024\\ 
	\multicolumn{11}{l}{1--10 years before billboard ban introduction} \\ 
	Coefficient &       0.0003&       0.0002&           0.0000&      -0.0001&      -0.0001&      -0.0003&      -0.0003&      -0.0005&      -0.0002&           0.0000\\
	Std. error          &       0.0003&       0.0002&       0.0003&       0.0003&       0.0003&       0.0004&       0.0004&       0.0008&       0.0005&       0.0004\\
	P-value    &       0.2237&       0.5257&       0.9321&       0.6103&       0.6608&        0.4230&        0.4680&       0.5786&       0.6686&       0.9518\\
	\hline
	Soc-dem. &   & X &   & X & X & X & X & X & X & X \\  
	Policy &   &    & X & X & X & X & X & X & X & X \\  
	Interacted &   &    &   &   & X &   & X &   &   &   \\    
	\hline
	Obs.          &     1,170,804&     1,170,804&     1,153,149&     1,155,675&     1,098,297&     1,144,859&     1,099,310&      421,618&      754,529&      974,241\\ 
	\hline\hline
	\end{tabular}
	\caption*{\footnotesize \textit{Note: This table reports year-specific ATETs of tobacco billboard bans on the smoking rate, averaged over multiple pre- and posttreatment periods, estimated using the staggered DiD approach by \textcite{Callaway21}. The first three panels present posttreatment ATETs averaged over one, three, and five years following policy implementation, including the year of adoption (event time $=$ 0). The final three panels report pretreatment ATETs, i.e., placebo effects, for the year immediately preceding implementation, and for windows spanning over five and ten pretreatment years, beginning from the first pretreatment year. The specifications are as follows: (1) no covariates, (2) gender and age covariates, (3) sales and smoking ban covariates, (4) gender, age, sales ban, and smoking ban covariates, (5) all these covariates interacted, (6) never-treated units as a control group, (7) never-treated units as a control group (all covariates interacted), (8) smoking status reconstructed retrospectively for up to 5 years, (9) up to 10 years, and (10) up to 15 years. Specifications (1)--(5) and (8)--(10) use not-yet-treated units as the control group. The standard errors are clustered at the individual level.}}
\end{table}

\FloatBarrier

\newpage

\begin{figure}[htbp!]
	\centering
	\subfigure[Staggered DiD (-5 years)]{\includegraphics[width=0.45\textwidth]{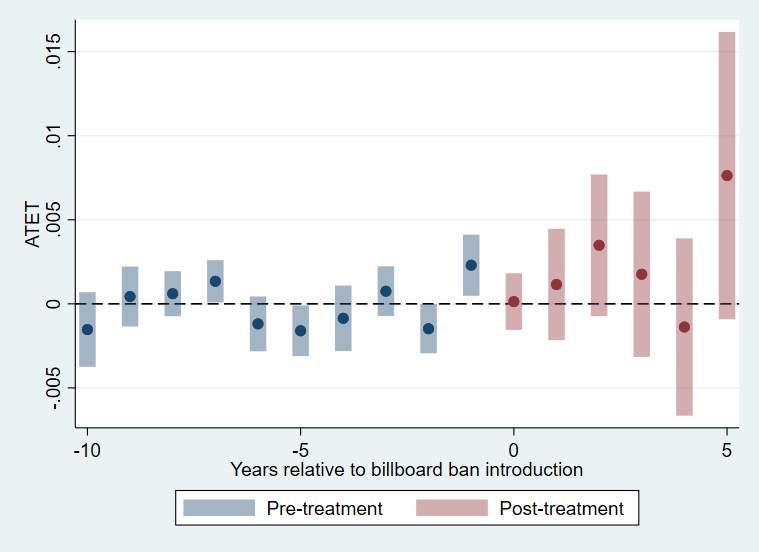}}
	\subfigure[IFEct (-5 years)]{\includegraphics[width=0.45\textwidth]{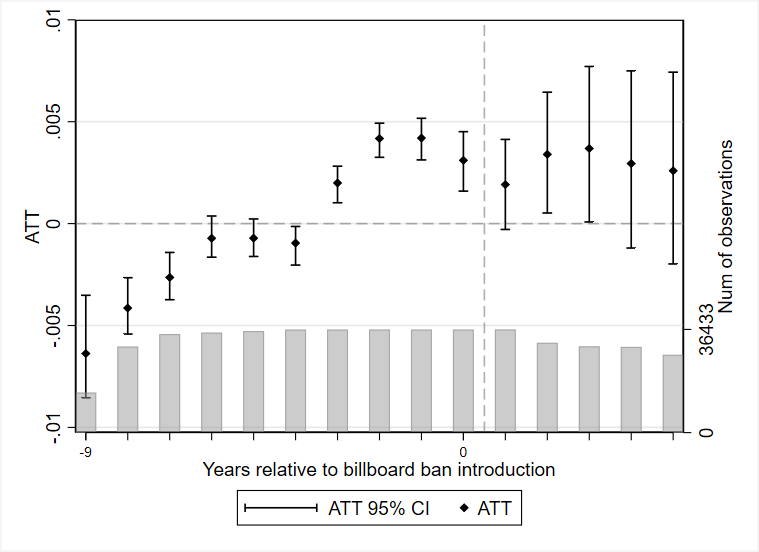}}
	\subfigure[Staggered DiD (-6 years)]{\includegraphics[width=0.45\textwidth]{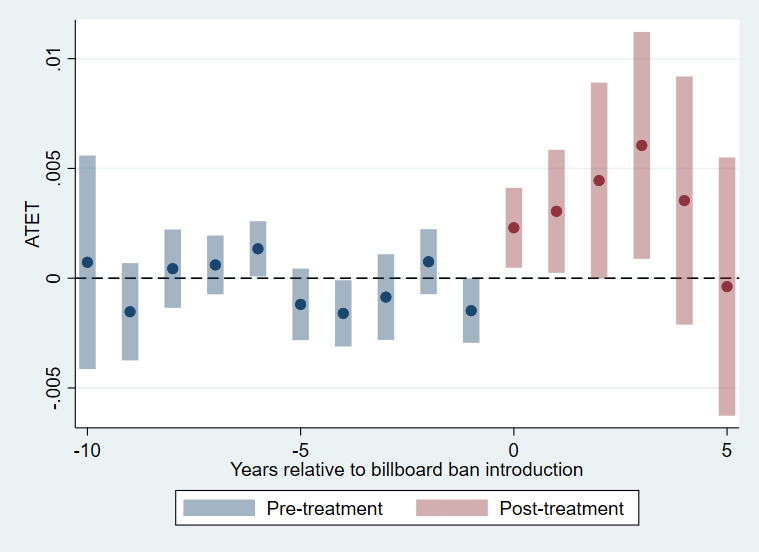}}
	\subfigure[IFEct (-6 years)]{\includegraphics[width=0.45\textwidth]{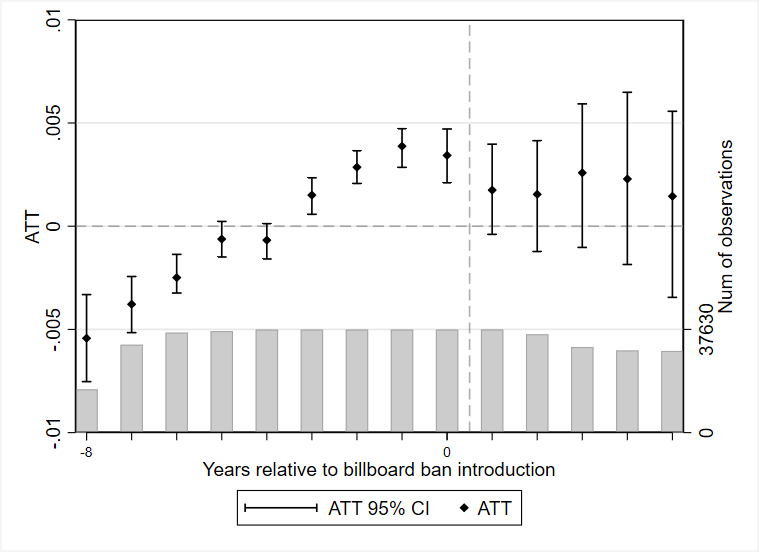}}
	\subfigure[Staggered DiD (-7 years)]{\includegraphics[width=0.45\textwidth]{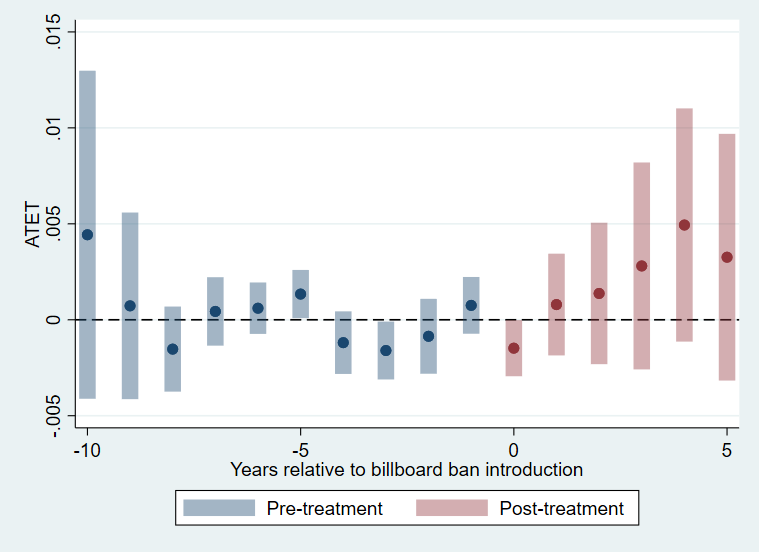}}
	\subfigure[IFEct (-7 years)]{\includegraphics[width=0.45\textwidth]{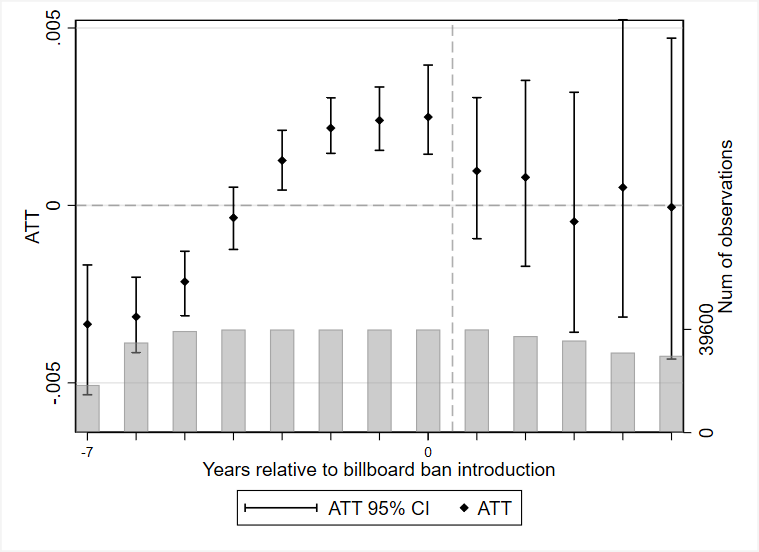}}
	\caption{Placebo event-study estimates}
	\label{fig:bb:mainResPlacebo}
	\caption*{\footnotesize \textit{Note: This figure presents placebo analyses based on the event-study specification for both the staggered DiD and IFEct estimators \parencite{Callaway21, Liu24}. Placebo treatments are created by assigning treatment 5, 6, and 7 years before the actual adoption of billboard bans. Error bars represent 95\% confidence intervals.}}
\end{figure}

\FloatBarrier

\newpage

\begin{table}[htbp!]
	\caption{Average year-specific effects of tobacco billboard bans on the smoking rate, by gender and age} 
	\label{tab:bb:hetRes}
	\centering
	\scriptsize
	\begin{tabular}{p{.1\textwidth}p{.08\textwidth}p{.08\textwidth}p{.08\textwidth}p{.08\textwidth}p{.08\textwidth}p{.08\textwidth}} 
		& \multicolumn{2}{c}{Gender} & \multicolumn{4}{c}{Age} \\
		\cmidrule(lr){2-3} \cmidrule(lr){4-7}
		&       Men &     Women&      15--24&       25--44&   45--64&     65+\\
		\hline
		\multicolumn{7}{l}{Average year-specific posttreatment effect} \\
		\hline
		\multicolumn{7}{l}{0--1 years after billboard ban introduction} \\ 
		ATET  &       0.0039&      -0.0078&       0.0036&      -0.0066&        0.0050&      -0.0087\\
		Std. error    &       0.0027&       0.0023&       0.0069&       0.0034&       0.0035&       0.0042\\
		P-value   &       0.1558&       0.0005&       0.5972&       0.0498&       0.1559&       0.0405\\
		\multicolumn{7}{l}{0--3 years after billboard ban introduction} \\ 
		ATET  &       0.0011&      -0.0102&       0.0003&      -0.0084&       0.0011&      -0.0103\\
		Std. error    &       0.0033&       0.0027&       0.0087&       0.0042&       0.0044&       0.0049\\
		P-value&       0.7457&       0.0002&       0.9715&       0.0431&       0.7988&       0.0378\\
		\multicolumn{7}{l}{0--5 years after billboard ban introduction} \\ 
		ATET  &       0.0012&      -0.0089&       0.0035&      -0.0067&       0.0004&      -0.0105\\
		Std. error     &       0.0038&       0.0031&       0.0097&       0.0048&        0.0050&       0.0056\\
		P-value &       0.7415&       0.0034&       0.7171&       0.1627&       0.9433&       0.0603\\
		\hline
		\multicolumn{7}{l}{Average year-specific pretreatment effect} \\
		\hline
		\multicolumn{7}{l}{1 year before billboard ban introduction} \\ 
		Coefficient  &       0.0043&       -0.0020&       0.0073&      -0.0028&        0.0020&       0.0024\\
		Std. error  &       0.0019&       0.0016&       0.0051&       0.0023&       0.0024&        0.0030\\
		P-value &        0.0210&       0.2042&       0.1511&       0.2255&       0.3958&       0.4267\\
		\multicolumn{7}{l}{1--5 years before billboard ban introduction} \\ 
		Coefficient  &       0.0009&      -0.0008&       0.0022&      -0.0012&      -0.0001&       0.0013\\
		Std. error  &       0.0006&       0.0005&       0.0019&       0.0008&       0.0009&       0.0011\\
		P-value  &       0.1734&       0.1165&       0.2339&       0.1274&       0.8645&        0.2450\\
		\multicolumn{7}{l}{1--10 years before billboard ban introduction} \\ 
		Coefficient  &       0.0004&      -0.0006&       0.0012&      -0.0008&      -0.0002&       0.0008\\
		Std. error  &       0.0004&       0.0003&        0.0010&       0.0005&       0.0006&       0.0007\\
		P-value  &       0.2848&       0.0878&       0.2523&       0.1185&       0.7542&       0.2513\\
		\hline
		Obs.  &      523,544&      632,162&      172,938&      468,979&      366,316&      145,680\\
		\hline\hline
	\end{tabular}
	\caption*{\footnotesize \textit{Note: This table reports year-specific ATETs of tobacco billboard bans on the smoking rate, averaged over multiple pre- and posttreatment periods, estimated using the staggered DiD approach by \textcite{Callaway21}, by gender and age. The first three panels present posttreatment ATETs averaged over one, three, and five years following policy implementation, including the year of adoption (event time $=$ 0). The final three panels report pretreatment ATETs, i.e., placebo effects, for the year immediately preceding implementation, and for windows spanning over five and ten pretreatment years, beginning from the first pretreatment year. The main specification is applied, including all covariates and cluster-robust standard errors.}}
\end{table}

\FloatBarrier

\newpage

\begin{figure}[htbp!]
	\centering
	\subfigure[Men]{\includegraphics[width=0.45\textwidth]{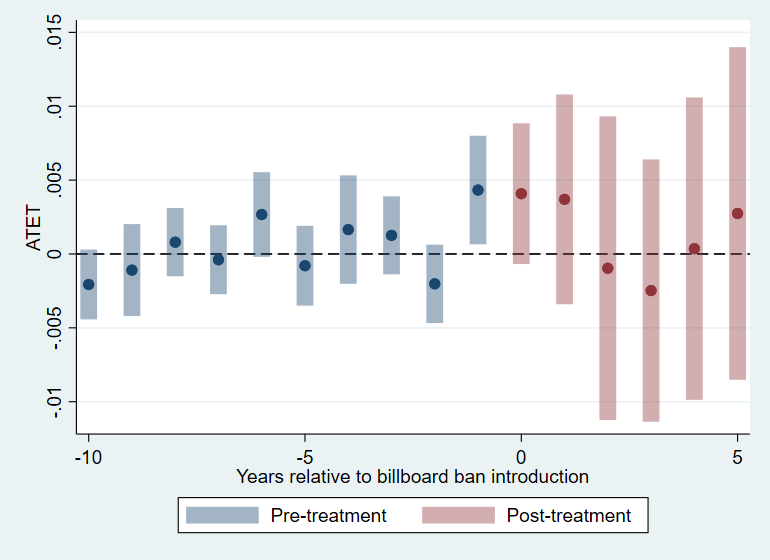}}
	\subfigure[Women]{\includegraphics[width=0.45\textwidth]{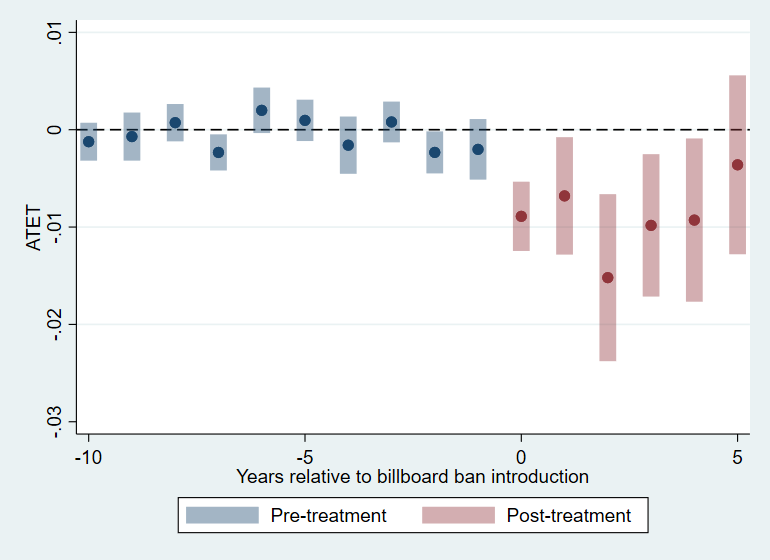}}
	\subfigure[15--24 years old]{\includegraphics[width=0.45\textwidth]{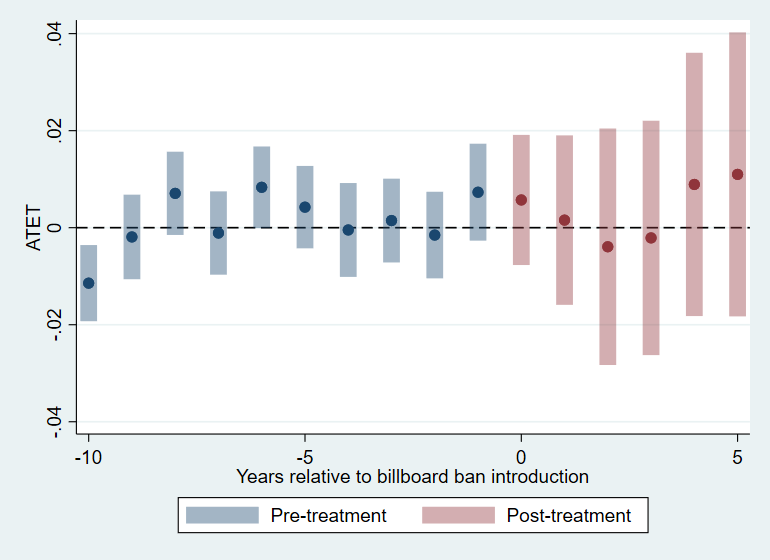}}
	\subfigure[25--44 years old]{\includegraphics[width=0.45\textwidth]{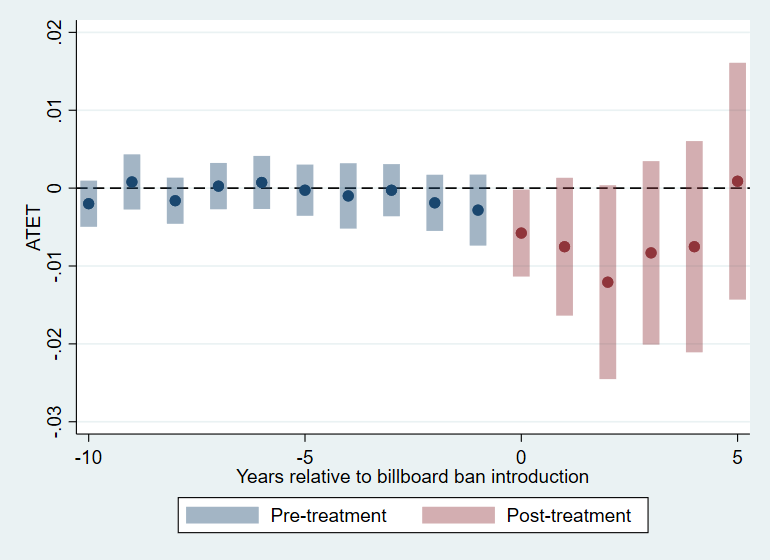}}
	\subfigure[45--64 years old]{\includegraphics[width=0.45\textwidth]{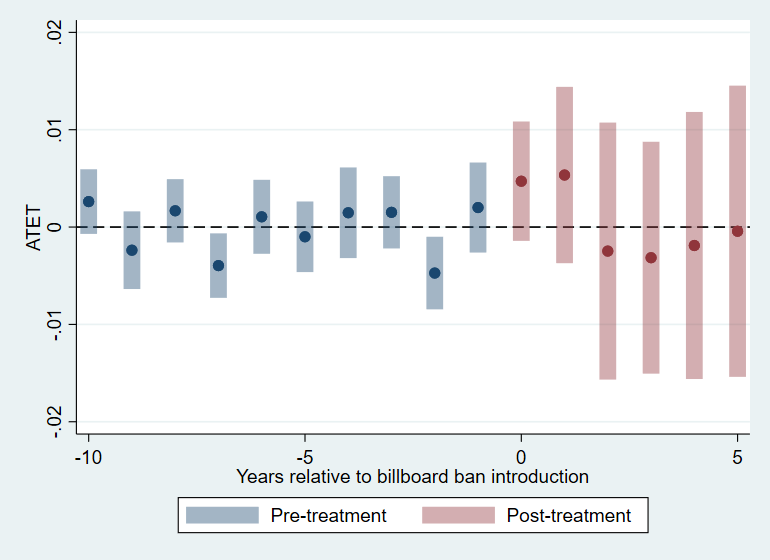}}
	\subfigure[65+ years old]{\includegraphics[width=0.45\textwidth]{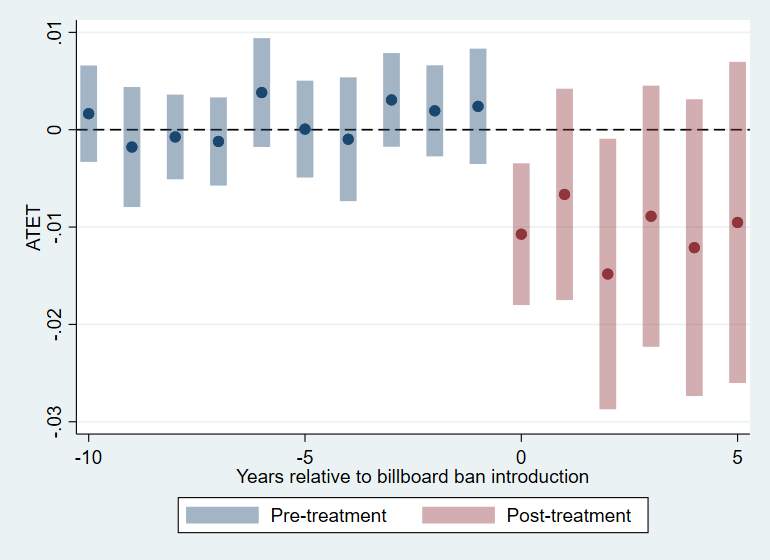}}
	\caption{Year-specific effects of tobacco billboard bans on the smoking rate, by gender and age}
	\label{fig:bb:hetRes}
	\caption*{\footnotesize \textit{Note: This figure presents year-specific ATETs of tobacco billboard bans on the smoking rate, estimated using the staggered DiD approach by \textcite{Callaway21}, by gender and age. Event time zero corresponds to the year of billboard ban implementation. The specification includes all covariates and cluster-robust standard errors. The point estimates are shown as dots, with 95\% confidence intervals represented by error bars.}}
\end{figure}

\FloatBarrier

\newpage

\begin{figure}[htbp!]
	\centering
	\subfigure[Men]{\includegraphics[width=0.45\textwidth]{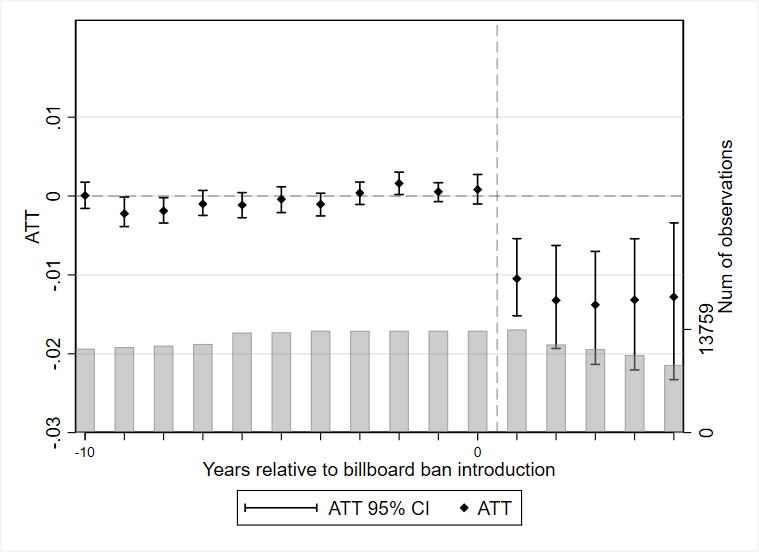}}
	\subfigure[Women]{\includegraphics[width=0.45\textwidth]{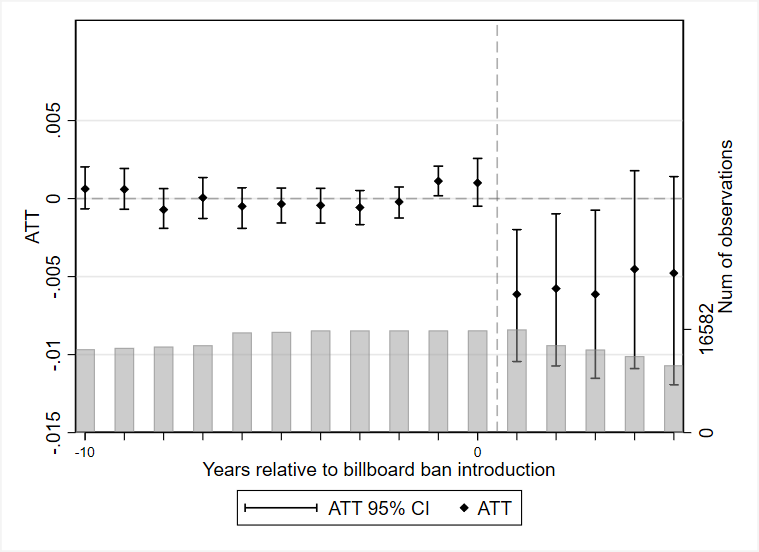}}
	\subfigure[15--24 years old]{\includegraphics[width=0.45\textwidth]{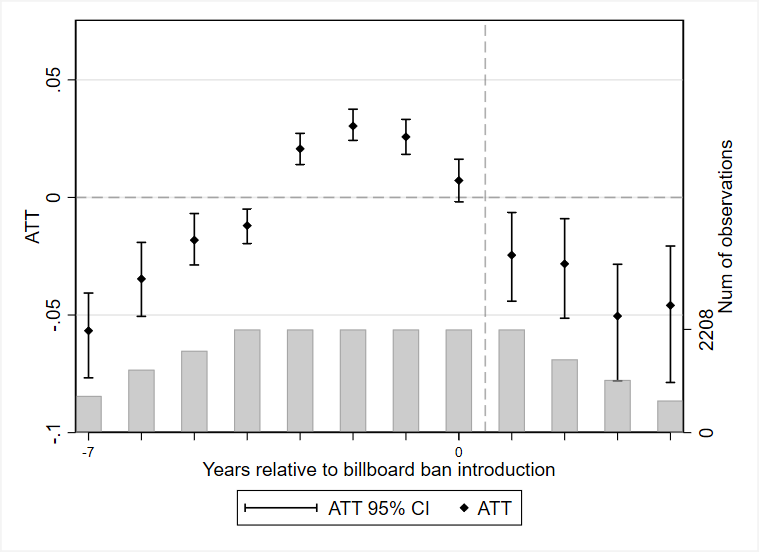}}
	\subfigure[25--44 years old]{\includegraphics[width=0.45\textwidth]{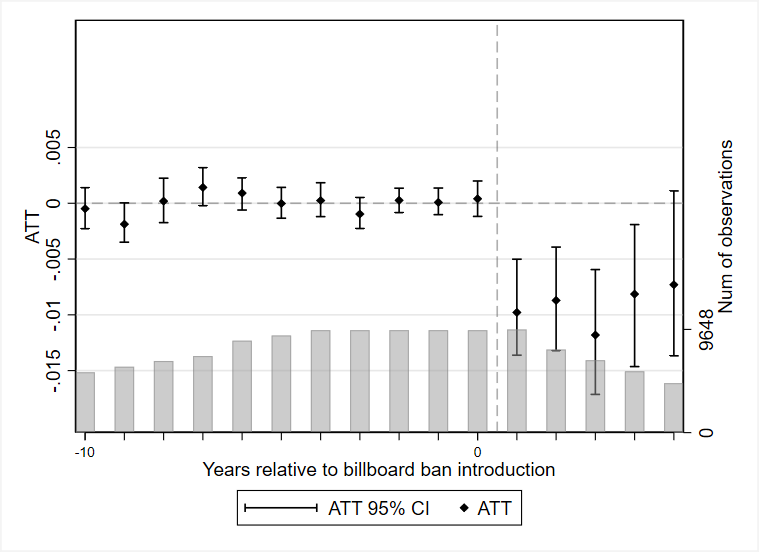}}
	\subfigure[45--64 years old]{\includegraphics[width=0.45\textwidth]{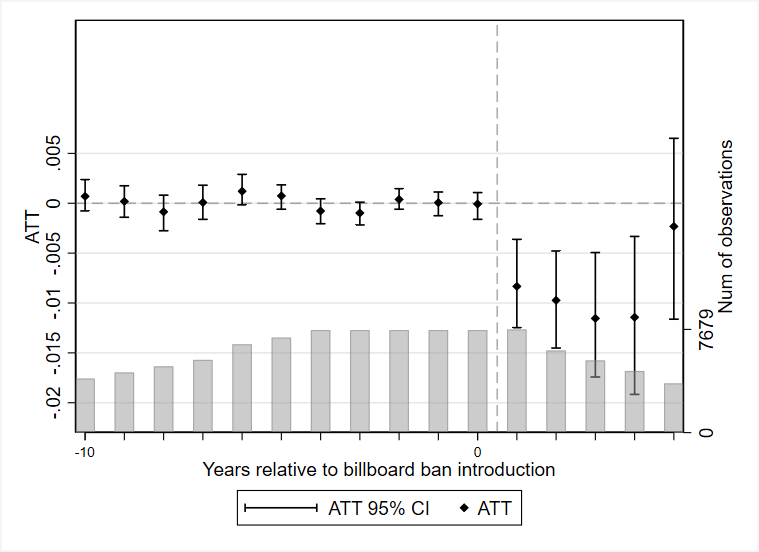}}
	\subfigure[65+ years old]{\includegraphics[width=0.45\textwidth]{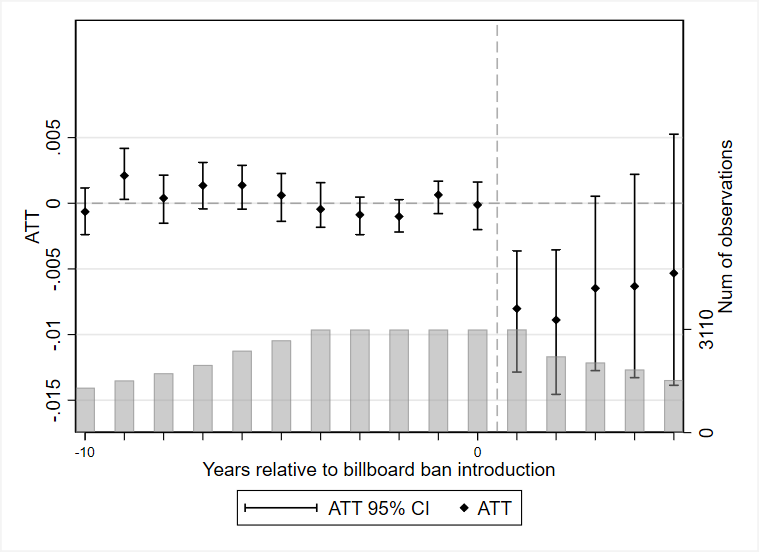}}
	\caption{Year-specific effects of tobacco billboard bans on the smoking rate, by gender and age}
	\label{fig:bb:ifeRes}
	\caption*{\footnotesize \textit{Note: This figure presents year-specific ATETs of tobacco billboard bans on the smoking rate, estimated using the IFEct estimator by \textcite{Liu24}, by gender and age. Event time zero corresponds to the year of billboard ban implementation. The specification includes all covariates and cluster-robust standard errors. The point estimates are shown as dots, with 95\% confidence intervals represented by error bars. The gray bars at the bottom show the number of included treated observations for the corresponding point estimate above.}}
\end{figure}

\FloatBarrier

\newpage

\begin{figure}[htbp!]
	\centering
	\includegraphics[width=0.6\textwidth]{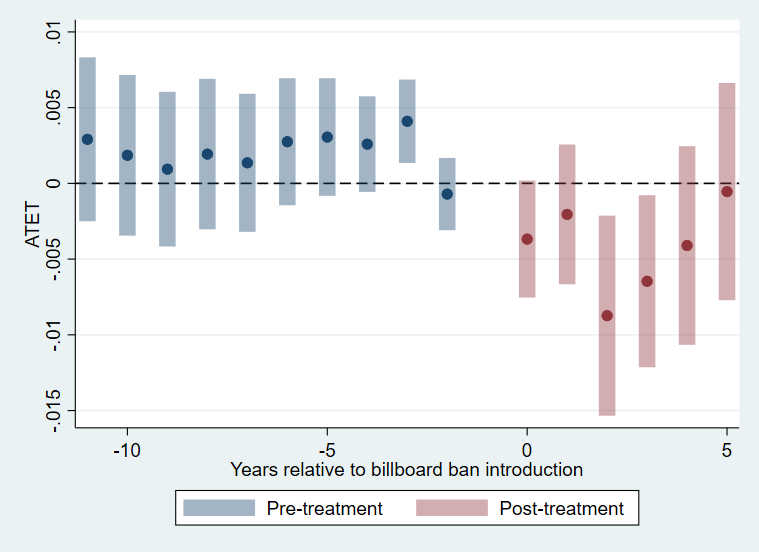}
	\caption{Year-specific effects of tobacco billboard bans on the smoking rate, using long gaps}
	\label{fig:bb:mainResNotYet_long}
	\caption*{\footnotesize \textit{Note: This figure presents year-specific ATETs of tobacco billboard bans on the smoking rate, estimated using the staggered DiD approach by \textcite{Callaway21}, with long gaps instead of short gaps. Event time zero corresponds to the year of billboard ban implementation. The specification includes all covariates and cluster-robust standard errors. The point estimates are shown as dots, with 95\% confidence intervals represented by error bars.}}
\end{figure}

\begin{figure}[htbp!]
	\centering
	\subfigure[Spillover dummy]{\includegraphics[width=0.45\textwidth]{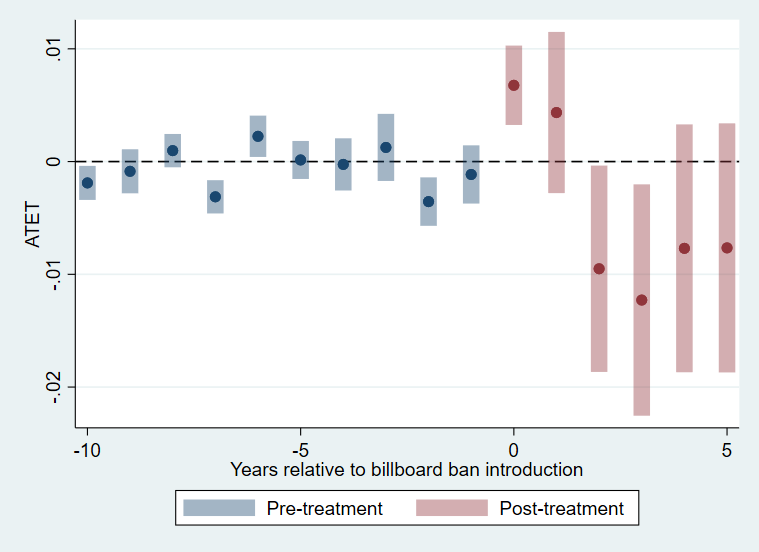}}
	\subfigure[Spillover dummy $\times$ commuter share]{\includegraphics[width=0.45\textwidth]{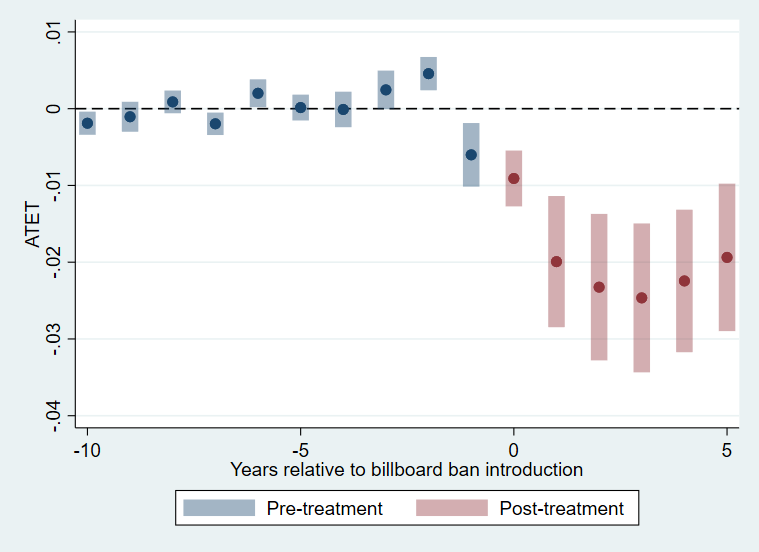}}
	\caption{Year-specific effects of tobacco billboard bans on the smoking rate, adjusting for spillover effects}
	\label{fig:bb:mainResSpill}
	\caption*{\footnotesize \textit{Note: This figure presents year-specific ATETs of tobacco billboard bans on the smoking rate, estimated using the staggered DiD approach by \textcite{Callaway21}, adjusting for spillover effects. Panel (a) augments the main specification with a dummy variable indicating whether a neighboring canton has introduced a billboard ban. Panel (b) further interacts this dummy with the share of commuters who travel outside their canton of residence. Event time zero corresponds to the year of billboard ban implementation. The specification includes all covariates and cluster-robust standard errors. The point estimates are shown as dots, with 95\% confidence intervals represented by error bars.}}
\end{figure}

\begin{figure}[htbp!]
	\centering
	\subfigure[No covariates]{\includegraphics[width=0.45\textwidth]{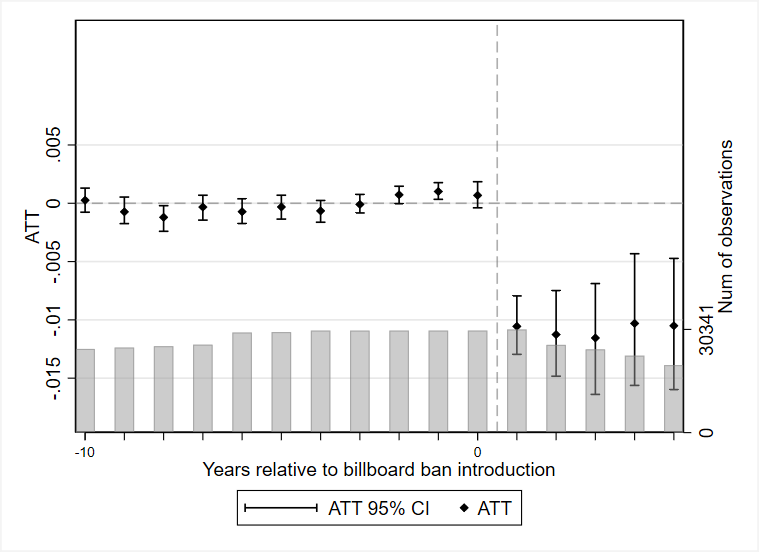}}
	\subfigure[Socio-demographic covariates]{\includegraphics[width=0.45\textwidth]{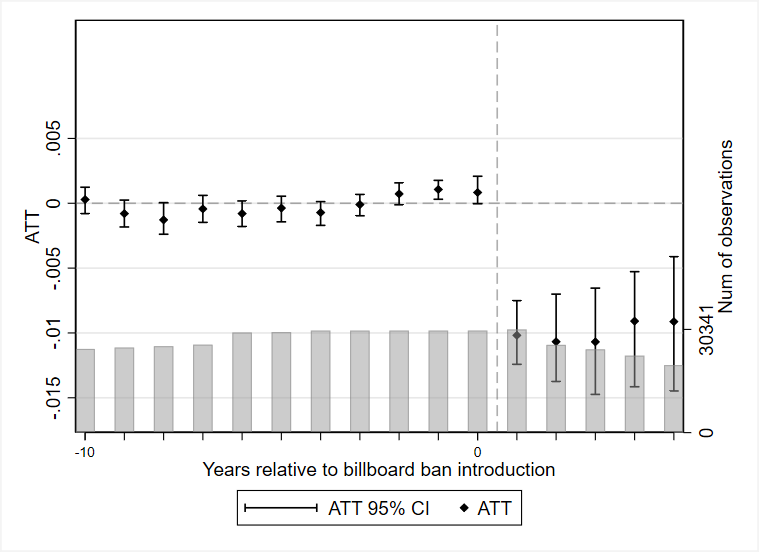}}
	\subfigure[Policy covariates]{\includegraphics[width=0.45\textwidth]{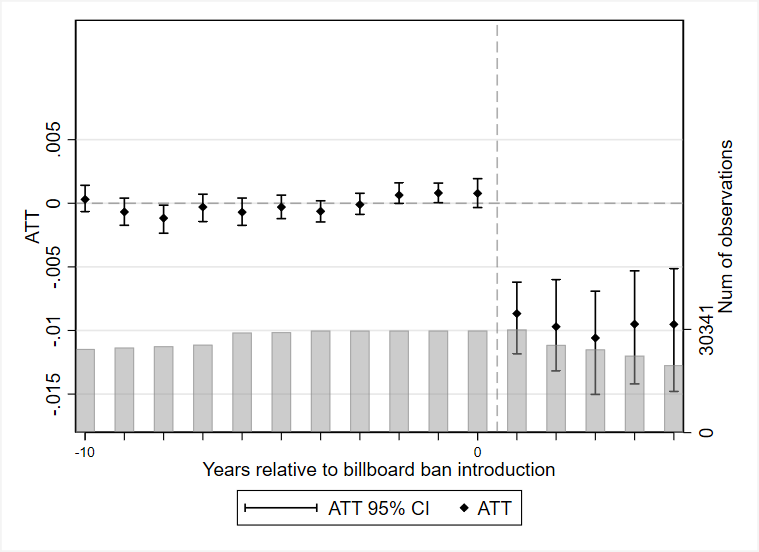}}
	\caption{Year-specific effects of tobacco billboard bans on the smoking rate, varying covariate compositions}
	\label{fig:bb:ifeResCovs}
	\caption*{\footnotesize \textit{Note: This figure presents year-specific ATETs of tobacco billboard bans on the smoking rate, estimated using the IFEct estimator by \textcite{Liu24}, varying the set of included covariates. Event time zero corresponds to the year of billboard ban implementation. The specification includes all covariates and cluster-robust standard errors. The point estimates are shown as dots, with 95\% confidence intervals represented by error bars. The gray bars at the bottom show the number of included treated observations for the corresponding point estimate above.}}
\end{figure}

\end{appendices}

\end{document}